\documentclass[preprint,12pt]{elsarticle}

\usepackage{amssymb}
\usepackage{url}            
\usepackage{booktabs}       
\usepackage{amsfonts}       
\usepackage{nicefrac}       
\usepackage{microtype}      
\usepackage{lipsum}
\usepackage{epsfig}

\usepackage[margin=1in]{geometry}
\usepackage{url}

\usepackage{graphicx,pstricks}
\usepackage{graphics}
\usepackage{amssymb}
\usepackage{amsmath, amsthm}
\usepackage{amsfonts}
\usepackage{algorithm}
\usepackage{booktabs,tabu,multirow}
\usepackage{subfig}
\usepackage{caption}
\usepackage{colortbl}
\usepackage{setspace}
\usepackage{xr}
\journal{Journal of the Mechanical Behavior of Biomedical Materials}

\begin{document}
\begin{frontmatter}

\title{Effect of Right Ventricular Outflow Tract Material Properties on Simulated Transcatheter Pulmonary Placement}

\author[1]{Jalaj Maheshwari\corref{cor1}}
\author[2,3,6]{Wensi Wu\corref{cor1}}
\author[1]{Christopher N. Zelonis}
\author[4]{Steve A. Maas}
\author[5]{Kyle Sunderland}
\author[1]{Yuval Barak-Corren}
\author[1]{Stephen Ching}
\author[1]{Patricia Sabin}
\author[5]{Andras Lasso}
\author[6]{Matthew J. Gillespie}
\author[4]{Jeffrey A. Weiss}
\author[1,6]{Matthew A. Jolley\corref{cor2}}

\cortext[cor1]{Authors contributed to the work equally.}
\cortext[cor2]{Corresponding author. }
\ead{jolleym@chop.edu}

\affiliation[1]{organization={Department of Anesthesiology and Critical Care Medicine, Children's Hospital of Philadelphia},
                 city={Philadelphia},
                 postcode={19104}, 
                 state={PA},
                country={USA}}
\affiliation[2]{organization={Cardiovascular Institute, Children's Hospital of Philadelphia},
                 city={Philadelphia},
                 postcode={19104}, 
                 state={PA},
                country={USA}}
\affiliation[3]{organization={Department of Mechanical Engineering and Applied Mechanics, University of Pennsylvania},
                 city={Philadelphia},
                 postcode={19104}, 
                 state={PA},
                country={USA}}
\affiliation[4]{organization={Scientific Computing and Imaging Institute, University of Utah},
                 city={Salt Lake City},
                 state={UT},
                country={USA}}
\affiliation[5]{organization={Laboratory for Percutaneous Surgery, Queen’s University},
                 city={Kingston},
                 state={ON},
                country={Canada}}
\affiliation[6]{organization={Division of Pediatric Cardiology, Children's Hospital of Philadelphia},
                 city={Philadelphia},
                 postcode={19104}, 
                 state={PA},
                country={USA}}

\begin{abstract}
Finite element (FE) simulations emulating transcatheter pulmonary valve (TPV) system deployment in patient-specific right ventricular outflow tracts (RVOT) assume material properties for the RVOT and adjacent tissues. Sensitivity of the deployment to variation in RVOT material properties is unknown. Moreover, the effect of a transannular patch stiffness and location on simulated TPV deployment has not been explored. A sensitivity analysis on the material properties of a patient-specific RVOT during TPV deployment, modeled as an uncoupled HGO material,  was conducted using FEBioUncertainSCI. Further, the effects of a transannular patch during TPV deployment were analyzed by considering two patch locations and four patch stiffnesses. Visualization of results and quantification were performed using custom metrics implemented in SlicerHeart and FEBio. Sensitivity analysis revealed that the shear modulus of the ground matrix ($c$), fiber modulus ($k_1$), fiber dispersion (\(\kappa\)), and fiber mean orientation angle (\(\gamma\)) had the greatest effect on 95th \%ile stress, whereas only $c$ had the greatest effect on 95th \%ile Lagrangian strain. First-order sensitivity indices contributed the greatest to the total-order sensitivity indices. Simulations using a transannular patch revealed that peak stress and strain were dependent on patch location and the underlying vessel geometry, with increasing patch stiffness modulating the magnitude of local stress concentrations. As stiffness of the patch increased, greater stress was observed at the interface connecting the patch to the RVOT, and stress in the patch itself increased while strain decreased. The total enclosed volume by the TPV device remained unchanged across all simulated patch cases. This study highlights that while uncertainties in tissue material properties and patch locations may influence functional outcomes, FE simulations provide a reliable framework for evaluating these outcomes in TPVR.

\end{abstract}


\end{frontmatter}

\section{ABBREVIATIONS}
    
\begin{description}
    
    \item[2D]
    Two Dimensional
    \item[3D]
    Three Dimensional
    \item[CTA]
    Computed Tomography Angiography
    \item[DICOM]
    Digital Imaging and Communications in Medicine
    \item[FE]
    Finite Element
    \item[HGO]
    Holzapfel Gasser Ogden
    \item[RVOT]
    Right Ventricular Outflow Tract
    \item[STL]
    Stereolithography
    \item[ToF]
    Tetralogy of Fallot
    \item[TPV]
    Transcatheter Pulmonary Valve
    \item[PVR]
    Pulmonary Valve Replacement
    \item[TPVR]
    Transcatheter Pulmonary Valve Replacement
    \item[TAVI]
    Transcatheter Aortic Valve Implantation
    \item[PCE]
    Polynomial Chaos Expansion
\end{description}

\section{INTRODUCTION} 
Pulmonary insufficiency is a consequence of surgical repair in Tetralogy of Fallot (ToF), leading to right ventricular dilation, ventricular arrhythmias, and heart failure. Pulmonary valve replacement (PVR), traditionally with a surgically implanted bioprosthetic valve, is often required to prevent many of these long-term complications \cite{geva_repaired_2011, lee_long-term_2020, bokma_improved_2023, geva_long-term_2024}. In the small subset of ToF patients who had a cylindrical RVOT conduit placed as part of their initial repair, a balloon-expandable transcatheter PVR (TPVR) has been used without the need for cardiopulmonary bypass \cite{biernacka_transcatheter_2015, demkow_percutaneous_2014, ghawi_transcatheter_2012, kenny_percutaneous_2011, obyrne_trends_2015, valente_contemporary_2014}. However, the vast majority of patients with ToF undergo surgical repair using a transpulmonary annulus patch, resulting in a dilated and heterogeneous shape of the native RVOT.  As a result, balloon-expandable transcatheter valves cannot be safely used in over 75 percent of ToF patients in need of PVR \cite{gillespie_1-year_2023, gartenberg_transcatheter_2023, jolley_toward_2019, kenny_current_2017, patel_transcatheter_2022, schievano_variations_2007, capelli_patient-specific_2010}. 

The limited applicability of balloon-expandable valves within these complex outflow tracts has stimulated the development of self-expanding valve platforms designed to conform to a variety of native RVOT shapes and sizes \cite{benson_three-year_2020, zahn_first_2018, jin_five-year_2024}. These self-expanding TPVR systems can conform to a wide range of heterogeneous native RVOT shapes, but precise matching of the optimal TPV to an individual patient is required \cite{benson_three-year_2020, schoonbeek_implantation_2016, gillespie_patient_2017}. RVOT anatomies do not correspond to a single characteristic shape; prior morphological analyses have grouped RVOTs into distinct geometric categories \cite{schievano_variations_2007}. This shape variation can strongly influence device positioning and procedural feasibility. As the number of available devices for TPVR continues to increase, along with the number of potential candidates and different RVOT anatomies, the ability to screen patients efficiently and choose an appropriate device becomes critical \cite{gillespie_patient_2017, jolley_toward_2019, bergersen_harmony_2017, mcelhinney_transcatheter_2024}. 
    
Current screening techniques to identify suitable candidates are labor-intensive and continue to rely primarily on 2D measurements from CT reconstructions of the RVOT and occasional implantation of actual devices into 3D-printed models to visually assess device fit \cite{mcelhinney_transcatheter_2024}. More recently, image-based preprocedural planning tools have been developed to assess patient eligibility and landing-zone adequacy in dysfunctional RVOTs. These include patient-specific digital and 3D-printed models \cite{pluchinotta_3-dimensional_2020} and semi-automated pipelines that derive the geometry of the TPVI landing zone from 4D-CT imaging \cite{sturla_planning_2025}. While several studies have explored transcatheter aortic valve implantation (TAVI) \cite{auricchio_simulation_2014, morganti_simulation_2014, ovcharenko_modeling_2016, sturla_impact_2016, bosi_population-specific_2018}, computational simulation studies examining TPVR are sparse \cite{capelli_patient-specific_2010, bosi_patientspecific_2015, donahue_finite_2022, donahue_finite_2024}. While additional finite element (FE) simulations to inform matching of devices to an image-derived patient-specific anatomy are emerging \cite{zelonis_integrated_2025, astori_fe-based_2025}, all simulations assume material properties for the RVOT within the mediastinum based on very limited data \cite{cabrera_mechanical_2013, jia_experimental_2017}. Further, the material properties of a surgically altered right ventricular outflow tract could vary from those of normal tissue. For example, the effect of patch material in the RVOT, which may alter tissue local stiffness and compliance at the time of initial repair on simulated TPV deployment, has not yet been investigated. Unfortunately, it is not currently possible to extract patient-specific material properties of the RVOT and pulmonary artery, and the sensitivity of the simulated deployment of self-expanding TPV to variations in RVOT tissue properties, and the effect of incorporation of heterogeneous materials into the reconstructed RVOT remains unknown. As such, there is a critical need to understand the sensitivity of emerging FE simulations of TPVR to both variation of RVOT material properties and the heterogeneity of material properties throughout the RVOT.

In the present work, we performed a two-stage sensitivity analysis using our open-source simulation framework implemented within SlicerHeart \cite{lasso_slicerheart_2022} and FEBio \cite{maas_febio_2012} to better understand how uncertainties in tissue material parameters affect the simulated results of TPV deployment. First, we assessed the sensitivity of RVOT stress and strain distributions following TPV deployment to uncertainty in tissue properties, assuming uniform properties throughout the RVOT. Subsequently, we investigated the influence of nonuniform tissue properties introduced by the presence of a transannular patch. In particular, we examined how variations in patch stiffness relative to native tissue and its anatomical placement affect the simulated RVOT biomechanics of TPV deployment. We performed both a traditional parameter exploration as well as a polynomial chaos expansion (PCE)-based uncertainty quantification analysis using UncertainSCI \cite{burk_efficient_2020} to quantify these effects.

\section{METHODS}
\subsection{Procuring RVOT Model and Segmentation}
Patients with a diagnosis of ToF who underwent TPVR and for whom computer tomography angiography (CTA) of the RVOT prior to TPVR had been previously acquired were identified from an existing institutional database. A patient with a typical geometry for the RVOT was chosen as an example.  The Institutional Review Board at the Children’s Hospital of Philadelphia approved this study.

CTA images were acquired on a dual-source scanner (Siemens Healthcare, Forchheim, Germany) using a retrospective ECG-gating technique (2 x 128 x 0.6-mm slice collimation). Low-osmolar iodinated contrast (Iohexol, OmnipaqueTM, 350 mg/mL, GE Healthcare Inc.) was injected via peripheral intravenous access with a dose of 2 mL/kg (up to 100 mL) for the patient. These retrospectively EKG-gated acquired images were typically reconstructed choosing the 30\% phase for systole and 90\% for diastole.

CTA images were then imported into 3D Slicer (www.slicer.org) in Digital Imaging and Communication in Medicine (DICOM) format. CT images of the TPV25 device \cite{medtronic_tpv_2021} (Medtronic, Minneapolis, MN, USA) were acquired on a CT scanner and segmented to create templates for mesh creation  3D CAD software (Fusion 360, Autodesk, San Francisco, CA, USA). The dimensions of the TPV device were also confirmed using digital calipers. Segmentation of the RVOT used in this study was created using a deep learning-based module in 3D Slicer as used in previous research \cite{zelonis_integrated_2025, diaz-pinto_monai_2024}.

\subsection{Computational Model Setup}
The inner surface of the RVOT vessel was exported as a shell mesh from 3D Slicer as a stereolithography (.STL) file. This .STL file was imported into Blender (V4.4), and the RVOT surface was remeshed as a quad-dominant shell mesh with an average mesh size of 1.5 mm to reduce the mesh density. The quad-meshed surface was exported from Blender in a .ply format and imported into FEBio. To create a solid mesh for the RVOT, the shell mesh was then extruded outward, normal to each surface element to have a thickness of 1.5 mm  \cite{donahue_finite_2024, vanderveken_mechano-biological_2020}. 

To capture the anisotropic and hyperelastic material behavior and account for the fiber orientations in the vessel, the RVOT was assigned an uncoupled Holzapfel-Gasser-Ogden (HGO) constitutive material model \cite{gasser_hyperelastic_2006, maas_uncoupled_2025}. The deviatoric and volumetric behavior for this material model is controlled by the strain-energy function:

\begin{equation}\Psi_{r} = \tilde{\Psi}_{r} \left( \tilde{C} \right) + U(J),\end{equation}
where,
\begin{equation}\tilde{\Psi}_{r} = \frac{c}{2} \left( \tilde{ I_{1}} - 3 \right) + \frac{k_1}{2k_2} \sum_{\alpha=1}^{2} \left( \exp \left( k_2 \langle \tilde{E_{\alpha}} \rangle ^2 \right) - 1 \right),\end{equation}
and the volumetric strain energy function is:
\begin{equation}U(J) =  \frac{k}{2} \left( \frac{J^2 - 1}{2} - \ln J \right).\end{equation}
Here, \(\tilde{C}\) is the right Cauchy-Green deformation tensor, \(\tilde{I}_1 = tr{ \tilde{C} }\), \(\tilde{I}_{4 \alpha} = a_{\alpha r} \cdot\tilde{C} \cdot a_{\alpha r}\), and 
\(\alpha = 1, 2\), and \begin{math}
E_{\alpha} = \kappa \left( \tilde{I}_1 - 3 \right) + ( 1 - 3 \kappa ) \left( \tilde{I}_{4\alpha} - 1 \right)
\end{math}
represents the fiber strain. \(J\) is the determinant of the deformation tensor or the local volume ratio. \(a_{\alpha r}\) (\(\alpha = 1, 2\)) represents the vectors along which the two fiber families are oriented. The input parameters for the HGO material included the material density (\(\rho\)), shear modulus of the ground matrix (\(c\)), fiber modulus (\(k_1\)), fiber exponential coefficient (\(k_2\)), fiber mean orientation angle (\(\gamma\)), fiber dispersion (\(\kappa\)), and bulk modulus (\(k\)). The HGO model parameters were defined as per previously published work \cite{donahue_finite_2022} and are included in Table 1. Local material axes were assigned to all RVOT solid mesh elements such that the local X-axis was positioned along the maximum curvature or circumference of the vessel, the local Y-axis was positioned along the minimum curvature or longitudinal direction, and the local Z-axis was positioned normal to the vessel surface.

The TPV device was modeled using 1-dimensional beam elements. The beams had a circular cross-sectional area with a diameter of 0.375 mm. Input parameters for beam elements included the material density per unit length (\(\rho\)), the cross-sectional area of the beam (\(A\)), the shear-corrected cross-sectional areas (\(A_1\), \(A_2\)), Young's modulus (\(E\)), shear modulus (\(G\)), and the second moments of inertia (\(I_1\), \(I_2\)). Cross-sectional area (\( \pi r^{2} \)) and moments of inertia (\(\frac{1}{4} \pi r^4\)) corresponding to a circle were used. Material properties for Nitinol \cite{vemury_behaviour_2019} were used to model the TPV device and are included in Table 1.

To uniformly compress the TPV25 device before expanding it in the RVOT, a curved tube covering the TPV25 device with its center along the vessel centerline was used. The tube radius was 28 mm, which was just large enough to contain the TPV device completely without having it penetrate the tube. The tube was split into two parts, allowing for a staged release of the TPV device: first expansion at the distal half, and second expansion at the proximal half of the device. The tube surface was modeled using a triangular shell mesh with a thickness of 2 mm. A neo-Hookean material model was used for the tubes \cite{maas_neo-hookean_2025}, the inputs for which were the material density (\(\rho\)), Young's modulus (\(E\)), and Poisson's ratio ($\nu$). Parameter values for the tube are also included in Table \ref{tab:material_parameters}.

\begin{table}[!ht]
    \centering
\caption{Material parameters for simulation components.}
\label{tab:material_parameters}
    \begin{tabular}{cc} \hline
         \textbf{Material Parameter}& \textbf{Value}\\ \hline 
         \multicolumn{2}{c}{\textbf{RVOT}}\\ \hline 
         $\rho$& 1.02$\times 10^{-6}$ kg/mm3\\ 
         $c$
& 200 kPa\\ 
         $k_1$
& 13480 kPa\\ 
         $k_2$
& 1.06\\  
         $\gamma$
& 18.85 deg\\ 
         $\kappa$
& 0.33\\  
         $k$& 1500 kPa\\ \hline 
         \multicolumn{2}{c}{\textbf{TPV25}}\\ \hline
 $\rho$
&6.5$\times 10^{-6}$ kg/mm$^3$\\
 $E$&4.0$\times 10^{7}$ kPa\\
 $G$&1.5$\times 10^{7}$ kPa\\
 $A = A_1 = A_2$&0.111045 mm$^2$\\
 $I_1 = I_2$&0.000970722\\\hline
 \multicolumn{2}{c}{\textbf{Tube}}\\\hline
 $\rho$&6.5$\times 10^{-6}$ kg/mm$^3$\\
 $E$&4.0$\times 10^{7}$ kPa\\
 $\nu$&0.33\\\hline
    \end{tabular}
    
\end{table}

\begin{figure}[!h]
    \centering
    \includegraphics[width=1.0\linewidth]{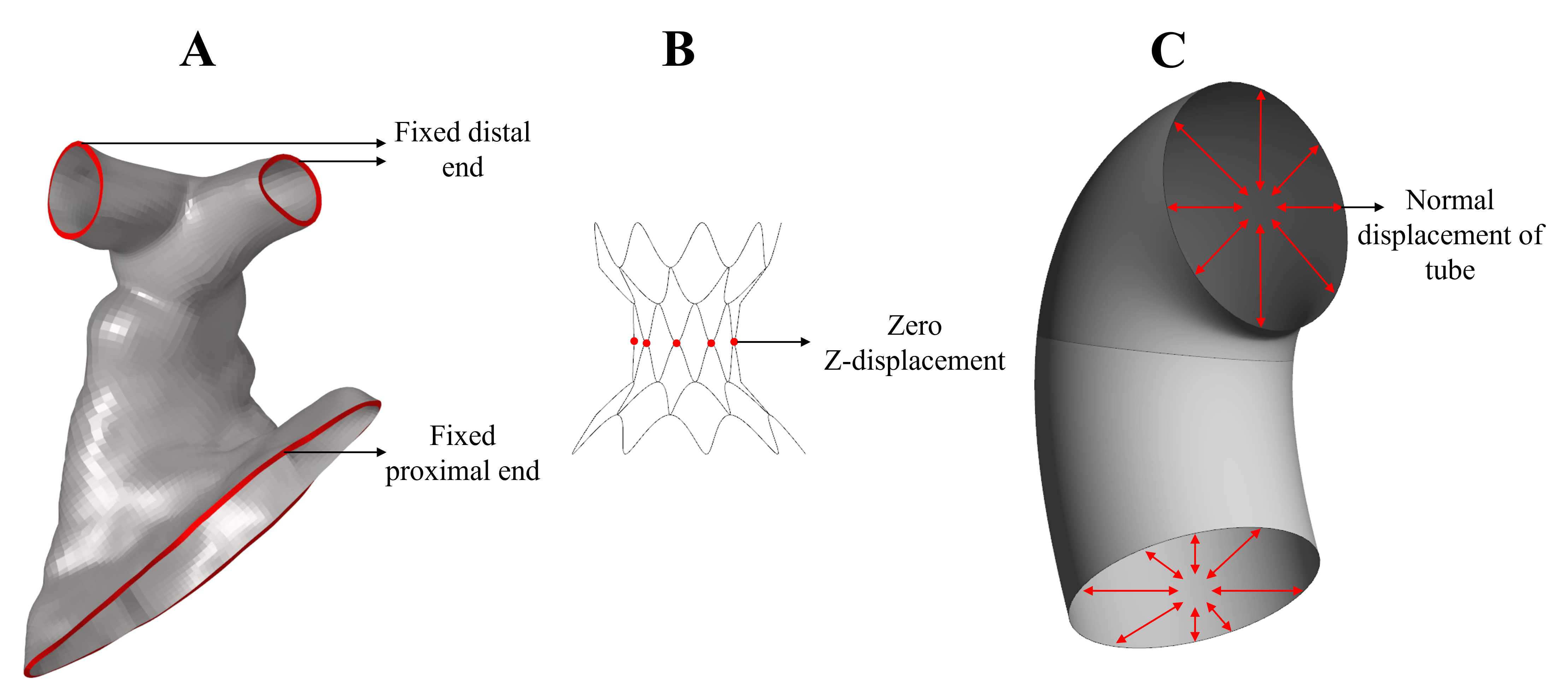}
    \caption{Boundary conditions for simulation components. (\textbf{A}) Distal and proximal ends of the vessel are fixed. (\textbf{B}) Central nodes of the TPV25 device and constrained to only move in the X-Y plane and are fixed along the vertical Z-axis. (\textbf{C}) The tubes surrounding the stent are provided a normal displacement to the surface of the tubes. This simulates a radial compression and expansion of the tube, and hence, the stent.}
    \label{fig:Boundary_Conditions}
\end{figure}

The distal and proximal ends of the RVOT were fixed. A zero displacement boundary condition was applied to the central nodes of the TPV25 device to prevent it from sliding in the Z direction or vertically along the vessel; the nodes were free to translate in the X-Y plane. The deployment of the TPV device in the RVOT was simulated in three stages: 1) compression of the device, 2) expansion of the device at the distal end, and 3) expansion of the device at the proximal end. The compression and expansion of the device were controlled by applying a normal displacement to the surrounding tube to mimic the self-expanding behavior of the TPV device. The boundary conditions for individual simulation components are depicted in Figure \ref{fig:Boundary_Conditions}. Edge-to-surface contact was defined between the TPV device and the tube for all stages, and between the TPV device and the RVOT vessel in the expansion stages. The edge-to-surface contact in FEBio expands on the sliding-facet-on-facet contact formulation, where the beam elements act as the primary surface instead of facets \cite{maas_sliding_2026}. It uses Gaussian quadrature to integrate the contact equations and uses a penalty-based method for constraint enforcement. This entire setup process, with the exception of the one dimensional beam elements stent, was also previously implemented in our emerging open-source framework \cite{zelonis_integrated_2025}. The simulation file for this setup has been uploaded to the FEBio model repository (\url{repo.febio.org/permalink/project/152}).

\subsection{Analysis Setup}
\subsubsection{Mesh convergence}
A mesh convergence analysis was conducted to determine the optimum mesh count for the RVOT. The RVOT wall was meshed with one through six layers of solid mesh. The intramural strain at the distal and proximal locations where the TPV device impinged the RVOT wall, and the 95th\%ile and 99th\%ile strains in the entire vessel when the stent expands completely were analyzed to determine the appropriate mesh density to use for subsequent analyses.

\subsubsection{Sensitivity and uncertainty analysis for RVOT material}
A sensitivity analysis was conducted to understand the effect of the HGO material model coefficients (\(c\), \(k_1\), \(k_2\), \(\gamma\), \(\kappa\)) on the TPV device deployment inside the RVOT. A Python subroutine called FEBioUncertainSCI, which interfaces UncertainSCI \cite{burk_efficient_2020} and the FEBio solver code \cite{maas_febio_2012} (\url{www.febio.org}), and which has been used in other sensitivity analyses of valve material properties \cite{wu_computational_2022, wu_effects_2023}, was used. For all HGO material parameters explored, baseline material properties from prior literature exploring RVOTs and self-expanding stents were chosen \cite{donahue_finite_2022}, which are also reported in Table \ref{tab:material_parameters}.

To find the range of values to use for each HGO material parameter for the sensitivity analysis, adult artery data were used to obtain the standard deviation value of the artery’s material properties from values reported for different arterial layers \cite{holzapfel_determination_2005}. Since the media layer is known to dominate the mechanical behavior of arteries \cite{holzapfel_new_2000, volokh_modeling_2011}, values for only the media layer were used. The standard deviation reported as a percentage of the mean for different arterial layers as per \cite{holzapfel_determination_2005} is shown in Table \ref{tab:HGO_material_table}A.

\begin{table}[!ht]
    \centering
    \caption{ (\textbf{A}) Standard deviation as a percentage of the mean for different arterial layers as per \cite{holzapfel_determination_2005}. (\textbf{B}) Mean and standard deviation of HGO material parameters used to calculate gamma distribution parameters. (\textbf{C}) Shape and scale parameters for gamma distributions for each HGO material parameter.}\label{tab:HGO_material_table}
\resizebox{\columnwidth}{!}{%
\begin{tabular}{cccccccccc}
\hline
\textbf{A} & \textbf{\begin{tabular}[c]{@{}c@{}}Arterial Layer\end{tabular}} & \textbf{Metric} & \textbf{$c$} & \textbf{$k_1$}              & \textbf{$k_2$}              & \textbf{\(\gamma\)}              & \textbf{\(\kappa\)}               & \textbf{epsilon}               & \textbf{Average}               \\ \hline
           & \multirow{3}{*}{Media}                                            & mean            & 1.27       & 21.6                     & 8.21                     & 20.61                       & 0.25                         & 0.05                           &                                \\
           &                                                                   & std             & 0.63       & 7.12                     & 3.27                     & 5.5                         & 0.09                         & 0.02                           &                                \\
           &                                                                   & percentage      & 49.61      & 32.96                    & 39.83                    & 26.69                       & 36                           & 48.89                          & 39                             \\    \hline\\ \hline
\textbf{B} & \multicolumn{3}{c}{\textbf{\begin{tabular}[c]{@{}c@{}}Material Parameter\end{tabular}}}        & \multicolumn{3}{c}{\textbf{\begin{tabular}[c]{@{}c@{}}Mean Value\end{tabular}}} & \multicolumn{3}{c}{\textbf{\begin{tabular}[c]{@{}c@{}}Standard Deviation (\%)\end{tabular}}} \\ \hline
           & \multicolumn{3}{c}{$c$}                                                                            & \multicolumn{3}{c}{200 kPa}                                                       & \multicolumn{3}{c}{49.61}                                                                      \\
           & \multicolumn{3}{c}{$k_1$}                                                                           & \multicolumn{3}{c}{13480 kPa}                                                     & \multicolumn{3}{c}{32.96}                                                                      \\
           & \multicolumn{3}{c}{$k_2$}                                                                           & \multicolumn{3}{c}{1.06}                                                          & \multicolumn{3}{c}{39.83}                                                                      \\
           & \multicolumn{3}{c}{\(\kappa\)}                                                                        & \multicolumn{3}{c}{0.0833}                                                        & \multicolumn{3}{c}{36}                                                                         \\
           & \multicolumn{3}{c}{\(\gamma\)}                                                                        & \multicolumn{3}{c}{18.85 deg}                                                     & \multicolumn{3}{c}{26.69}                                                                      \\ \hline\\ \hline
\textbf{C} & \multicolumn{4}{c}{\textbf{\begin{tabular}[c]{@{}c@{}}Gamma Distribution Parameters\end{tabular}}}                      & \textbf{$c$}               & \textbf{$k_1$}                 & \textbf{$k_2$}                  & \textbf{\(\gamma\)}                 & \textbf{\(\kappa\)}                 \\ \hline
           & \multicolumn{4}{c}{Shape (k)}                                                                                               & 4.0631                   & 9.2050                      & 6.3035                       & 14.0379                        & 7.7160                         \\
           & \multicolumn{4}{c}{Scale (\(\theta\))}                                                                                           & 49.2230                  & 1464.4154                   & 0.1682                       & 1.3428                         & 0.0108                \\ \hline        
\end{tabular}%
}
\end{table}
Using the mean values for parameters as those reported by \cite{donahue_finite_2022} and the standard deviation percentages from \cite{holzapfel_determination_2005}, a gamma distribution was determined for each material parameter to obtain the shape (k) and scale factor ($\theta$) to use as inputs for UncertainSCI’s gamma distribution function. Since \cite{donahue_finite_2022} used $\kappa$=1/3, which represents isotropic fiber dispersion, and using any standard deviation value with a gamma distribution would cause a violation of the constraints for $\kappa$ (0 $\le$ $\kappa$ $\le$ 1/3), the mean value for $\kappa$ from \cite{holzapfel_determination_2005} was used and scaled by the isotropic limit 1/3. Final mean values and standard deviation percentages for each HGO material parameter are shown in Table \ref{tab:HGO_material_table}B. Obtained values for shape and scale factor for each HGO material parameter for their respective gamma distributions using mean and standard deviation values from Table \ref{tab:HGO_material_table}B are shown in Table \ref{tab:HGO_material_table}C. 

To determine the most appropriate polynomial chaos expansion (PCE) order to use for the uncertainty analysis, a PCE convergence analysis was conducted. Three independent runs for PCE orders \(p\ = 1, 2, 3, 4\) were simulated and 1st Principal stress and Lagrangian strain were extracted and compared across orders. The normalized root mean square error (NRMSE), defined as the root mean square error normalized by the range of the output was used to compare metrics across orders:
\begin{equation}
\mathrm{NRMSE} =
\frac{
\sqrt{
\frac{1}{N}
\sum_{i=1}^{N}
\left(
y_i^{\mathrm{FE}} - y_i^{\mathrm{PCE}}
\right)^2
}
}{
\max(y^{\mathrm{FE}}) - \min(y^{\mathrm{FE}})
}
\end{equation}

Here, $N$ denotes the total number of samples used in the comparison. 
$y_i^{\mathrm{FE}}$ represents FE model output at the $i$‑th sample, and $y_i^{\mathrm{PCE}}$ denotes the corresponding prediction obtained from the PCE model. To obtain a dimensionless error metric, the root mean square error is normalized by the range of the FE output metrics used, $\max(y^{\mathrm{FE}}) - \min(y^{\mathrm{FE}})$. NRMSE for 1st Principal stress and strain, and the variation of first-order Sobol indices were used to select the final PCE order for the sensitivity analysis.

For the final chosen PCE order, all HGO material input parameters were varied simultaneously to generate the appropriate combinations of materials. Simulations were conducted for each of the combinations of HGO material parameter values. The 95th\%ile, 75th\%ile, and mean 1st principal stress and Lagrangian strain of the RVOT, and total- and first-order Sobol indices for each material parameter were extracted.

To understand the extent to which the stress-strain response varied within the uncertainty analysis, an equibiaxial tension test was simulated for a unit-dimensional cube element (1 mm x 1 mm x 1 mm). The uncoupled HGO material model \cite{gasser_hyperelastic_2006, maas_uncoupled_2025} was used with material parameter values from the PCE generated samples. The faces of the cube at X=0, Y=0, and Z=0 were constrained. Equal displacement was prescribed to the cube's faces at X=1 and Y=1, stretching both faces by 50\%. The Cauchy stress and elongation ratio along the tensioned directions, X and Y, were extracted for all PCE sample values.

\subsubsection{Transannular patch stiffness}
\begin{figure}[!h]
    \centering
    \includegraphics[width=0.7\linewidth]{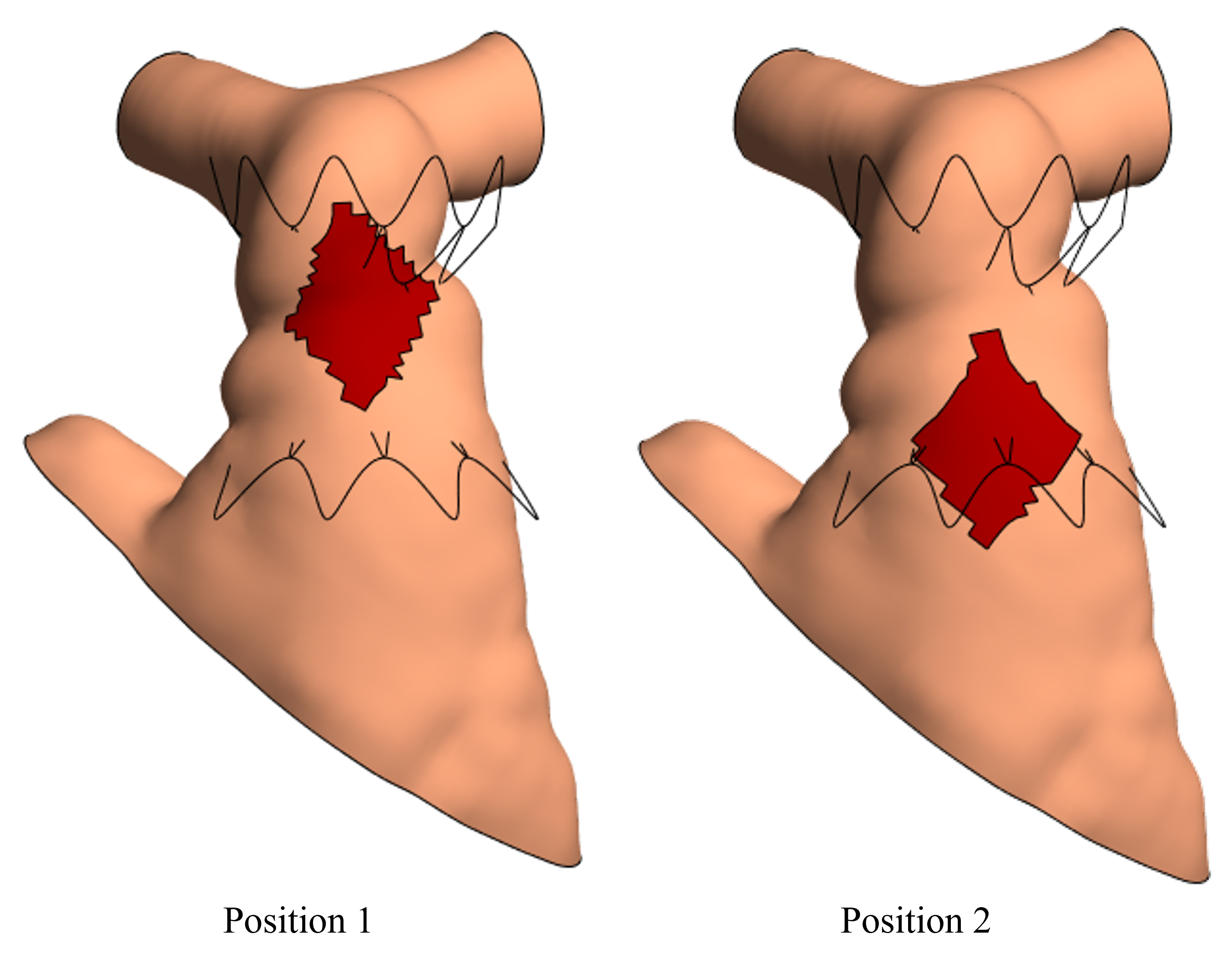}
    \caption{Transannular patch embedded in the RVOT vessel wall at two locations, a distal (location 1) and proximal (location 2) location with respect to the undeformed TPV device.}
    \label{fig:RVOT28_Transannular_patch_locations}
\end{figure}

To understand the effect of a transannular patch on the TPV device deployment in the RVOT, a diamond-shaped patch was embedded in the RVOT. Based on the largest size reported in prior literature \cite{rosenthal_aneurysms_1972}, the transannular patch measuring 30 mm x 20 mm along its longer and shorter diagonals was modeled and located at two locations where the TPV device interacts with the RVOT vessel wall, a distal location and a proximal location, shown in Figure \ref{fig:RVOT28_Transannular_patch_locations}. 

\begin{table}
    \centering
    \caption{Material parameter values for transannular patch conditions.}\label{tab:patch_materials}
\begin{tabular}{cccc}
\hline
\textbf{\begin{tabular}[c]{@{}c@{}}Transannular Patch \\ Stiffness Condition\end{tabular}} & \textbf{Density (kg/mm$^3$)}                 & \textbf{\begin{tabular}[c]{@{}c@{}}Poisson's \\ Ratio\end{tabular}} & \textbf{\begin{tabular}[c]{@{}c@{}}Young's\\ Modulus (kPa)\end{tabular}} \\ \hline
Stiffness 1                                                                      & \multirow{4}{*}{1.41$\times 10^{-6}$} & \multirow{4}{*}{0.495}                                              & 1.1 $\times 10^3$                                                     \\
Stiffness 2                                                                      &                                  &                                                                     & 2.2 $\times 10^3$                                                         \\
Stiffness 3                                                                      &                                  &                                                                     & 4.4 $\times 10^3$                                                          \\
Stiffness 4                                                                      &                                  &                                                                     & 8.8 $\times 10^3$       \\                              \hline                 
\end{tabular}
\end{table}

The transannular patch was modeled as an isotropic elastic material \cite{maas_isotropic_2025} with material properties based on prior literature \cite{m_measurement_2023, mc_dynamic_2015}, which have been highlighted in Table \ref{tab:patch_materials}. The stiffness of the patch was increased 2x, 4x, and 8x by scaling the Young’s modulus from values reported in prior literature, i.e., 1.1$\times 10^{3}$ kPa. The simulation files for the two transannular patch locations has been uploaded to the FEBio model repository (\url{repo.febio.org/permalink/project/152}). The maximum, 95th\%ile, 75th\%ile, and mean first principal stresses and Lagrangian strains were analyzed.

\section{RESULTS}

\subsection{Mesh Convergence Analysis}
\begin{figure}[!h]
    \centering
    \includegraphics[width=1.0\linewidth]{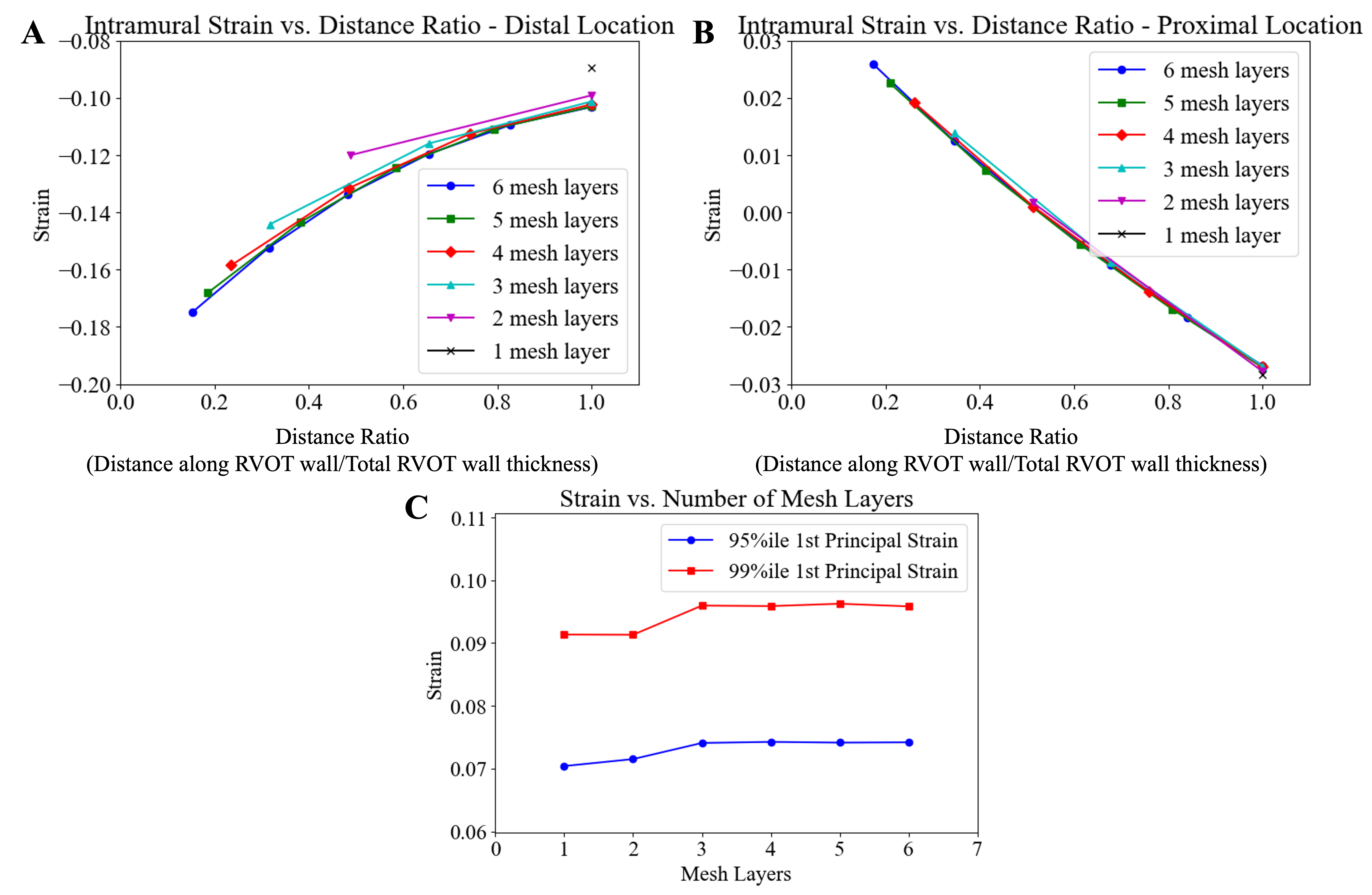}
    \caption{Mesh convergence analysis. (\textbf{A}) intramural strain vs. distance ratio at a distal location and (\textbf{B}) at a proximal location where the TPV device impinges the RVOT wall. (\textbf{C}) 95th\%ile and 99th\%ile strains in the entire RVOT at complete TPV device expansion. Based on the mesh convergence plots, the model containing 4 mesh layers in the RVOT wall was chosen for the subsequent simulations.}
    \label{fig:mesh_convergence}
\end{figure}

To determine the optimal mesh size for the RVOT, a mesh convergence analysis was conducted. Figure \ref{fig:mesh_convergence}A and \ref{fig:mesh_convergence}B show the intramural strain along the RVOT wall for different numbers of mesh layers (one through six) at the distal location and proximal location on the RVOT wall, respectively. In these figures, the x-axis represents the distance ratio or the distance along the vessel wall with respect to the total thickness of the RVOT wall (1.5 mm), and the y-axis represents the strain along the RVOT vessel wall. Figure \ref{fig:mesh_convergence}C represents the 95th and 99th\%ile 1st principal strains in the entire RVOT vessel for the different mesh layers across the vessel wall. The 95th and 99th\%ile strains were chosen to eliminate the effect of any `hotspots’ where the TPV device impinged on the vessel wall. From these figures, the intramural strains and 1st principal strains converged at 4 mesh layers through the RVOT vessel wall. Therefore, for all subsequent simulations and analyses, a 4-layered mesh RVOT wall FE model was used.

\subsection{Sensitivity and Uncertainty Analysis}
\subsubsection{PCE Convergence Analysis}
\begin{figure}[!h]
    \centering
    \includegraphics[width=1.0\linewidth]{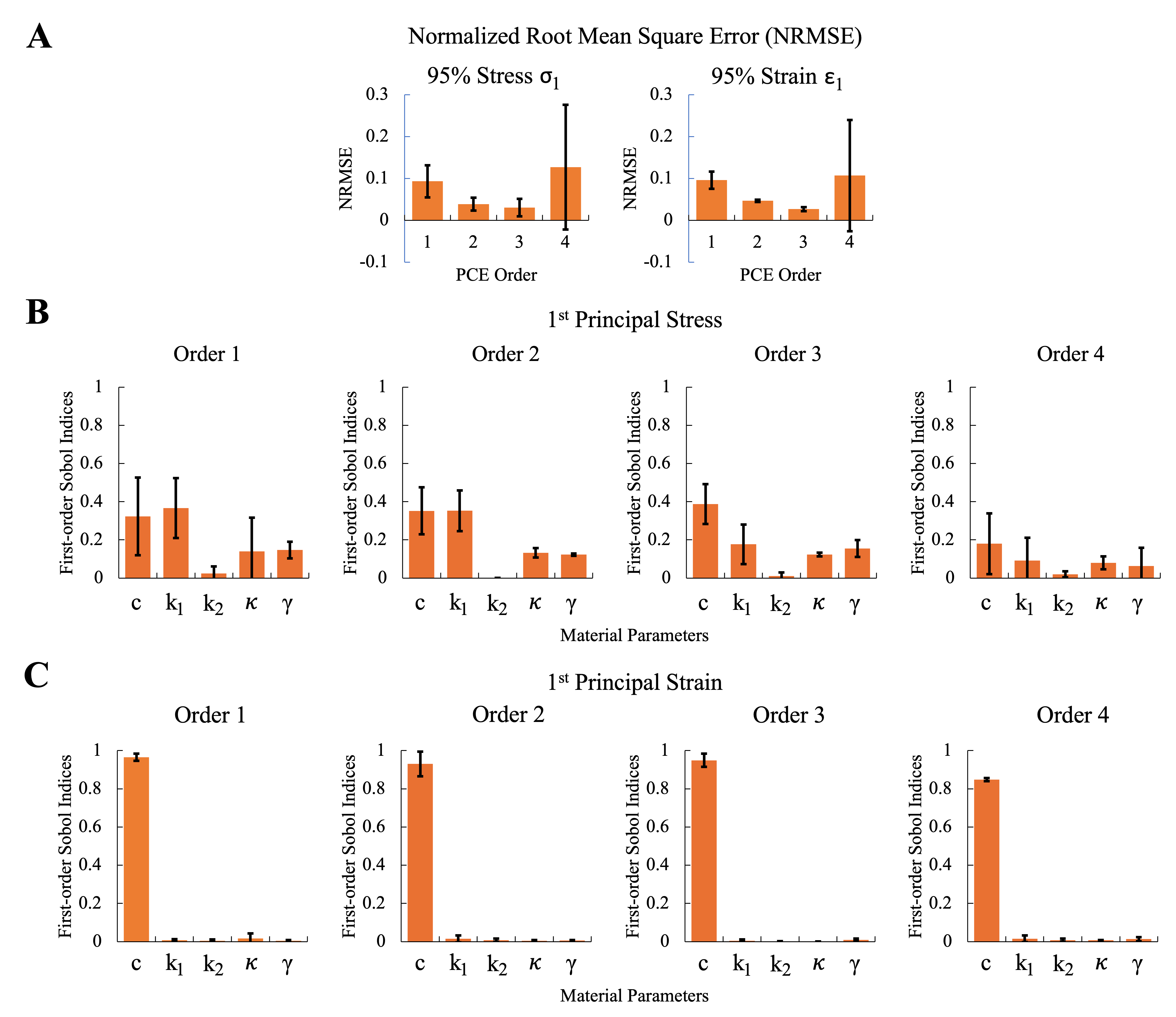}
    \caption{(\textbf{A}) Normalized root mean square error for 1st principal stress and 1st principal Lagrangian strain across three independent runs of PCE orders 1-4. Values depicted as mean and standard deviation for each order across the three runs. (\textbf{B}) First-order Sobol indices for each PCE order 1-4 across the three independent runs for 1st principal stress. (\textbf{C}) First-order Sobol indices for each PCE order 1-4 across the three independent runs for 1st principal Lagrangian strain.}
    \label{fig:PCE_convergence}
\end{figure}

Figure \ref{fig:PCE_convergence}A depicts the NRMSE for 95th\%ile 1st Principal stress and Lagrangian strain across PCE orders represented by their mean and standard deviation. Figure \ref{fig:PCE_convergence}B depicts the first-order Sobol indices for 1st Principal stress and Lagrangian strain for each HGO material parameter across runs. From these plots, PCE order 3 resulted in the least NRMSE and variation across sensitivity indices from the three independent runs. Therefore, a third-order polynomial chaos expansion function was used for the sensitivity analysis.

\subsubsection{Sensitivity Analysis}
A total of 66 simulations were generated for the sensitivity analysis which are included in the supplementary material as Table A1. Of these, 65 ran to completion successfully and 1 failed. The failed simulation had HGO material parameters that situated it on the lower extremity of the gamma distribution curves for each parameter, resulting in a highly distensible and unstable RVOT wall. Results were compiled for the remaining successful 65 simulations.

\begin{figure}
    \centering
    \includegraphics[width=1.0\linewidth]{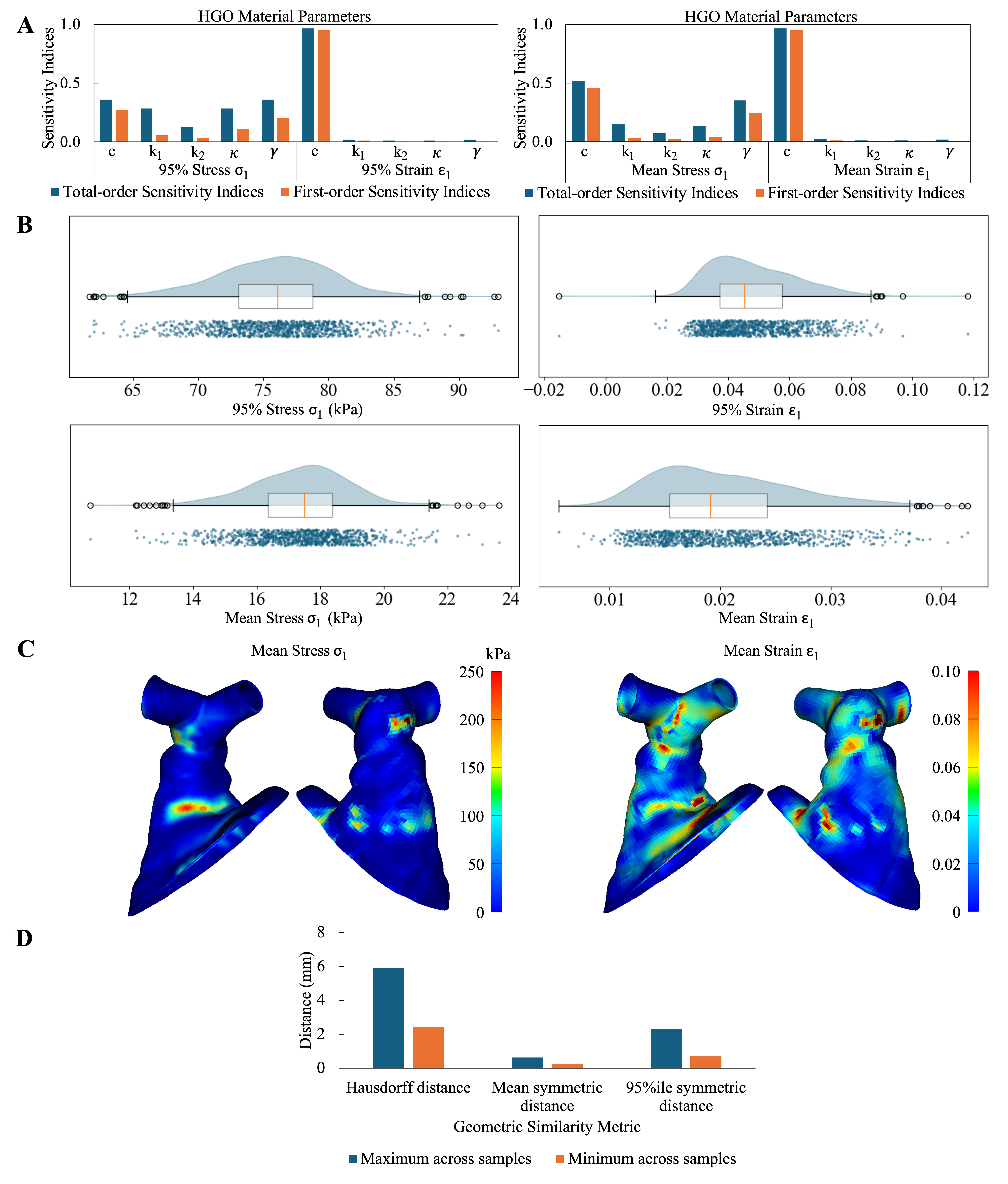}
    \caption{(\textbf{A}) Total and first-order sensitivity indices for the HGO material parameters for 95\%ile and mean 1st principal stress and Lagrangian strain in the RVOT. (\textbf{B}) Raincloud plot showing the distribution of 1000 sampling combinations of HGO material parameters for 95th\%ile and mean stress and strain in the RVOT vessel wall at maximum TPV device expansion. (\textbf{C}) Mean 1st principal stress and Lagrangian strain distribution across all simulations conducted for sensitivity analysis. (\textbf{D}) Geometric comparison metrics calculated across the 65 completed sensitivity simulations using a baseline simulation with vessel material parameters defined as per \cite{donahue_finite_2022, donahue_finite_2024}. Plot depicts the maximum and minimum values across all samples.}
    \label{fig:sensitivity_indices_3rdOrder}
\end{figure}

The total and first-order sensitivity indices at the 95th\%ile and mean stress and Lagrangian strain are shown in Figure \ref{fig:sensitivity_indices_3rdOrder}A. Within the sampling space, the shear modulus of the ground matrix ($c$), fiber modulus ($k_1$), fiber dispersion (\(\kappa\)), and fiber mean orientation angle (\(\gamma\)) had the greatest influence on 95th\%ile stress, while $c$ followed by \(\gamma\) had the greatest influence on 75th\%ile and mean stress in the RVOT vessel wall. The shear modulus of the ground matrix ($c$) also had the greatest influence on the 95th\%ile, 75th\%ile, and mean Lagrangian strain. Both total and first-order sensitivity indices followed similar patterns, with the first-order indices contributing the greatest to the total-order indices.

The input HGO material parameters and the corresponding output were used to train a PCE emulator as previously explored \cite{wu_computational_2022, wu_effects_2023}. 1000 sampling combinations of the HGO material parameters were queried from the PCE emulator. Figure \ref{fig:sensitivity_indices_3rdOrder}B depicts the raincloud plots showing the minimum, maximum, and median values, and the distribution of these 1000 combinations on the output metrics. High uncertainties were observed for all plotted metrics, i.e. 95th\%ile and mean stresses and strains in the entire RVOT model, indicating that the RVOT material is highly sensitive to the input material parameter values. Furthermore, the variation in each output metric was large, greater than 10\% of the difference in the minimum and maximum values.

Figure \ref{fig:sensitivity_indices_3rdOrder}C depicts the mean 1st principal stress and Lagrangian strain calculated across all simulations run for the sensitivity analysis to obtain potential `hot-spot' regions of high stress and strain. Regions of highest stress and strain, irrespective of the RVOT wall material parameters, were observed at the location of the proximal and distal ends of the stent, where the stent wireframe contacts the RVOT wall. Using the simulation with parameter values from \cite{donahue_finite_2022, donahue_finite_2024} as a baseline, similarity between the RVOT geometries at complete device expansion was determined by computing the Hausdorff distance, mean symmetric distance, and 95\%ile symmetric distance for the 65 completed simulations. Figure \ref{fig:sensitivity_indices_3rdOrder}D highlights the maximum and minimum metric values across the 65 simulations. The maximum and minimum Hausdorff distances were 19.5\% and 8.1\% of the perimeter-derived diameter at the narrowest region of the baseline simulation RVOT at complete stent expansion. These metric values highlight that while stress and strain values varied across the generated samples, stress and strain patterns and RVOT geometries were fairly similar on complete stent expansion.

\begin{figure}[!h]
    \centering
    \includegraphics[width=0.70\linewidth]{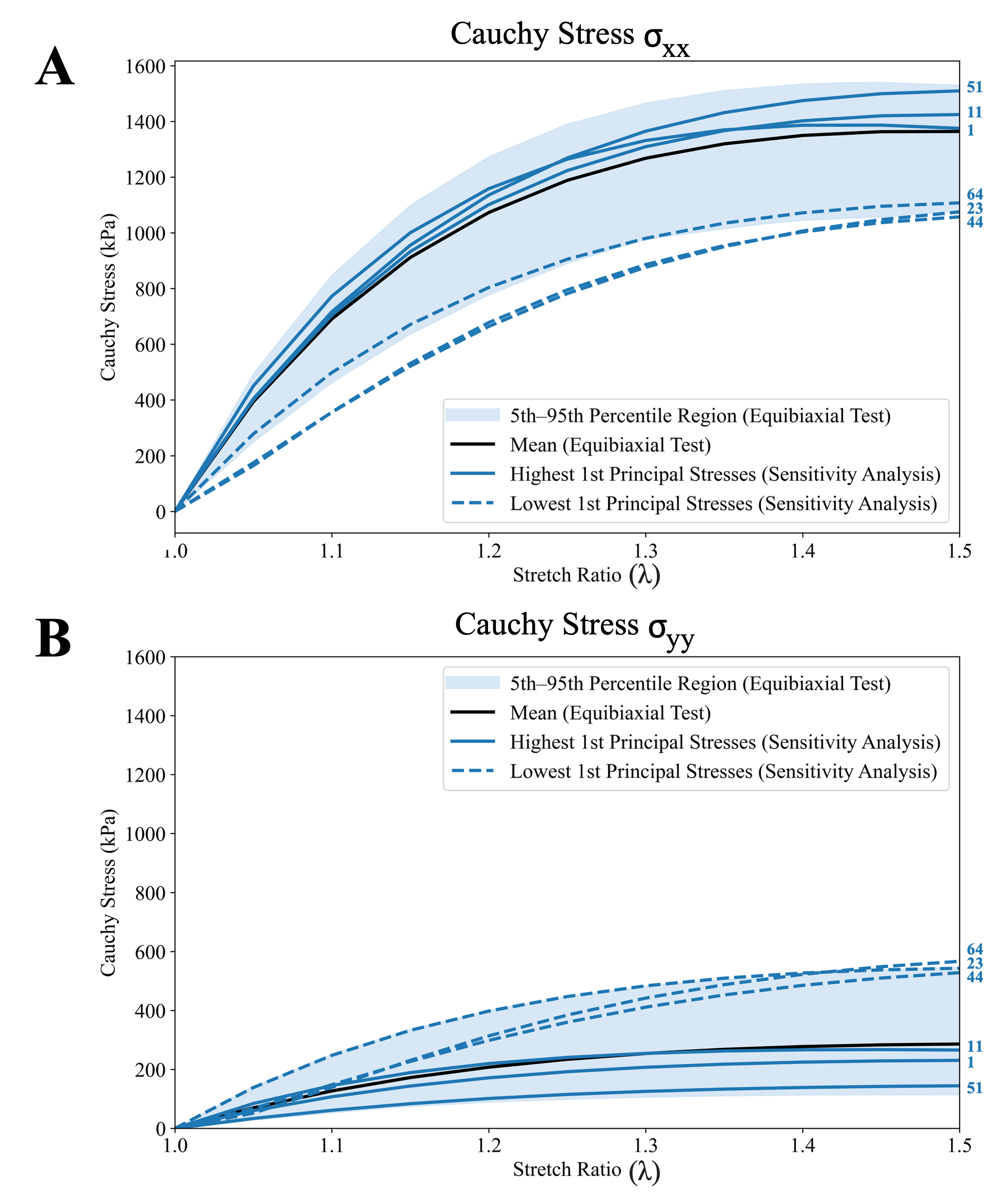}
    \caption{Cauchy stress vs. stretch ratio in the (\textbf{A}) X-direction and (\textbf{B}) Y-direction for the equibiaxial tension tests on a unit-dimensional cube. Mean curve (solid black) and 5th-95th\%ile region (blue shaded region) were computed across all 66 PCE samples generated for the sensitivity analysis. The three samples with highest 1st principal stress in the sensitivity analysis were plotted as solid blue curves. Similarly, the three samples with the lowest 1st principal stress in the sensitivity analysis were plotted as dashed blue curves.}
    \label{fig:equibiaxial_test_plot}
\end{figure}

The Cauchy stress in the X and Y directions versus stretch ratio for the equibiaxial tension test simulations are shown in Figures \ref{fig:equibiaxial_test_plot}A and \ref{fig:equibiaxial_test_plot}B. The mean 95\%ile stress curve and the 5th-95th\%ile region (represented by blue region) for this equibiaxial test were computed across all 66 PCE samples generated for the sensitivity analysis. From the RVOT sensitivity analysis, the three samples depicting the highest 95\%ile 1st principal stress (samples 1, 11, and 51) were plotted as solid blue curves. Similarly, the three samples depicting the lowest 95\%ile 1st principal stress (samples 23, 44, and 64) in the sensitivity analysis were plotted as dashed blue curves. From these plots and Table A1 (included in the supplementary materials), it can be observed that different combinations of HGO material parameter values result in similar curves.

\subsection{Transannular Patch Stiffness Analysis}

\begin{figure}[!h]
    \centering
    \includegraphics[width=0.60\linewidth]{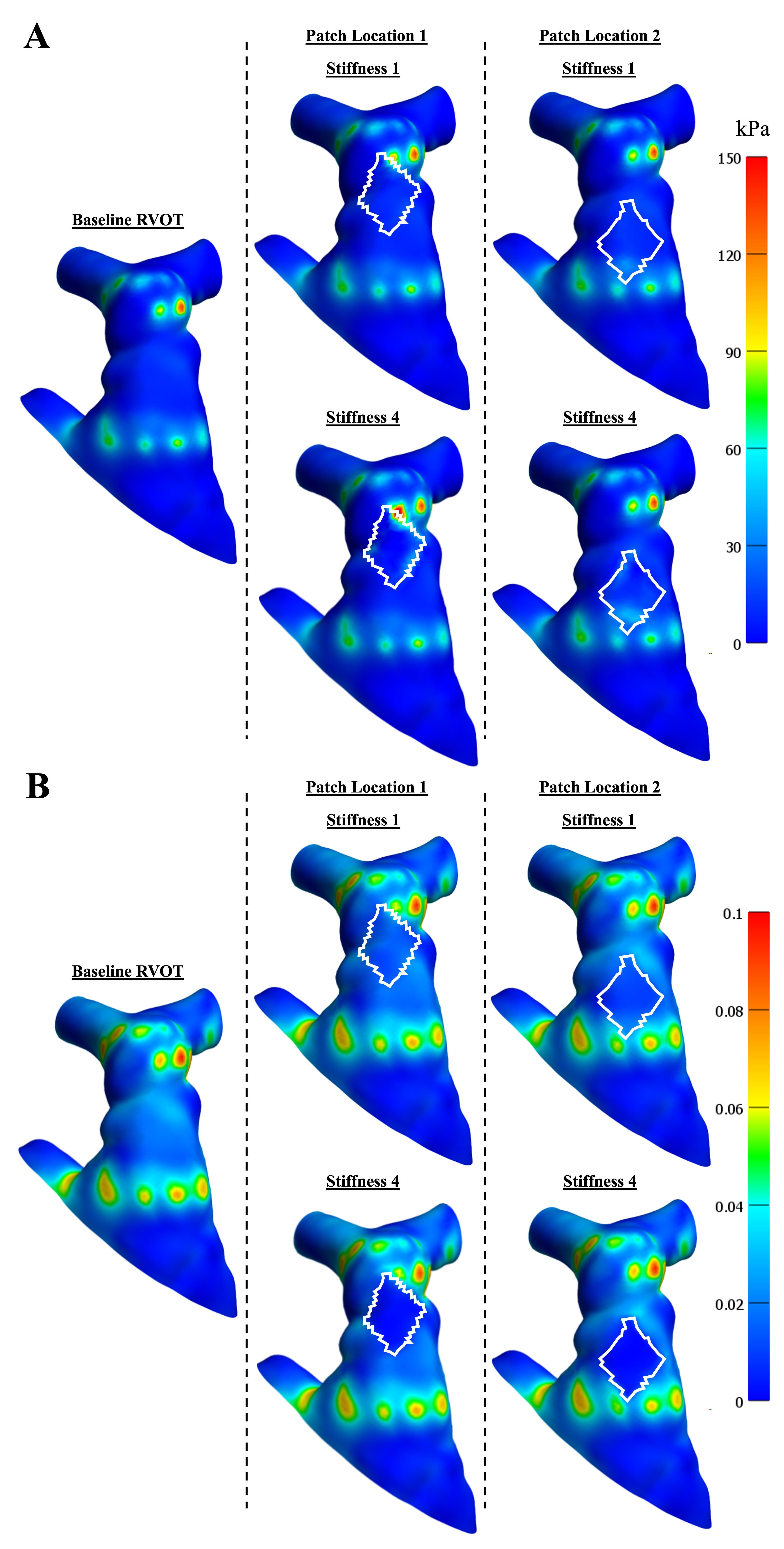}
    \caption{(\textbf{A}) 1st principal stress and (\textbf{B}) 1st principal Lagrangian strain distribution across baseline, patch location 1, and patch location 2 simulations for stiffness 1 and stiffness 4 conditions.}
    \label{fig:Patch_stress_strain_distribution}
\end{figure}

Figure \ref{fig:Patch_stress_strain_distribution} depicts the 1st principal stress and Lagrangian strain distribution for a baseline RVOT without a transannular patch, a patch in location 1, and a patch in location 2 for two stiffness values. Due to the presence of the patch, greater regions of stress are observed at the boundary where the patch attaches to the RVOT. Conversely, the strain observed in the patched region was minimal compared to that observed in the rest of the RVOT. Higher regions of stress and strain were also observed at the locations where the proximal and distal ends of the stent impinge on the RVOT wall. Additionally, as the stiffness of the patch increased, stress at the boundary of the patch and RVOT increased, while strain decreased.

\begin{figure}[!h]
    \centering
    \includegraphics[width=1\linewidth]{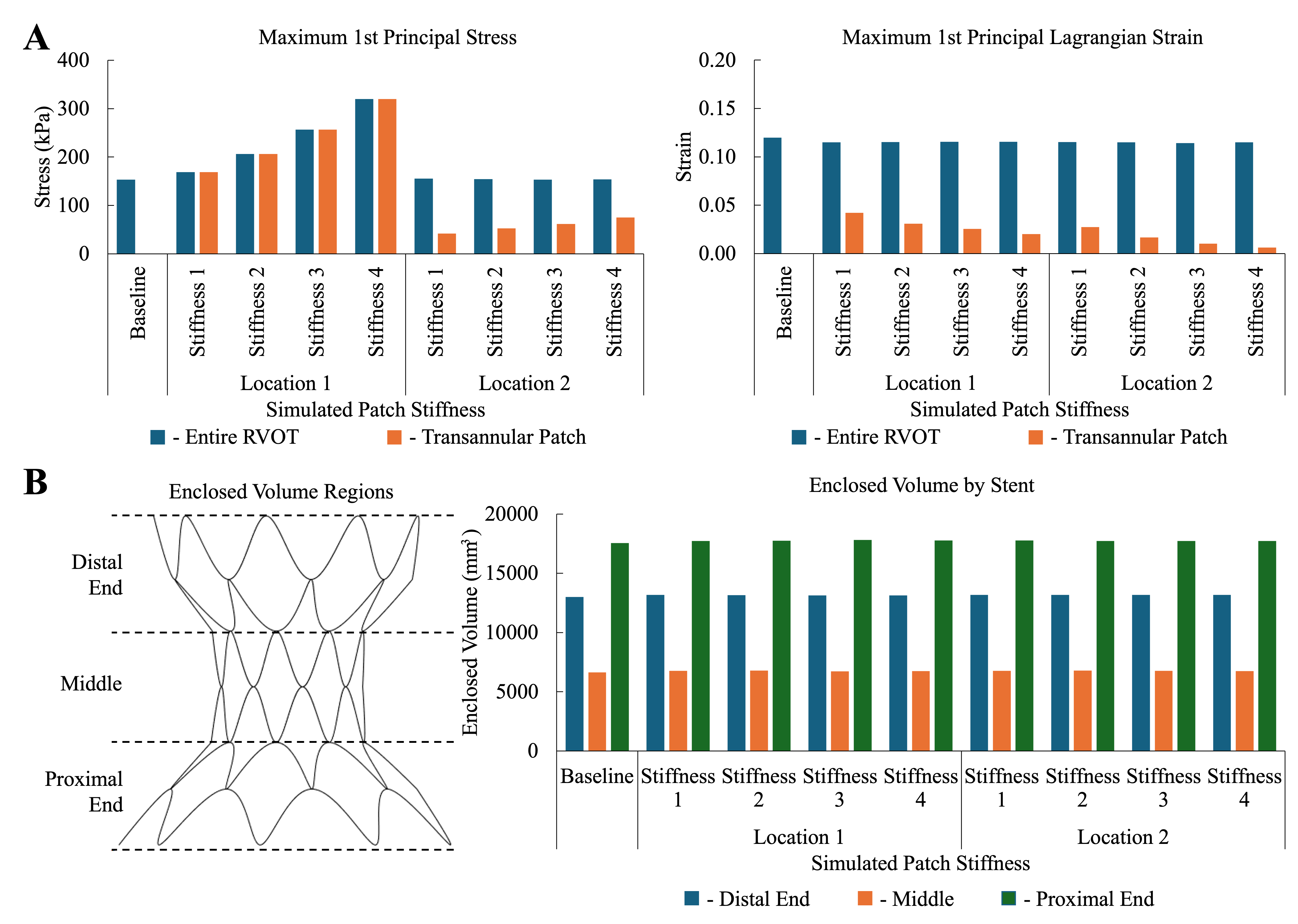}
    \caption{(\textbf{A}) 1st principal stress and 1st principal Lagrangian strain distribution across baseline, patch location 1, and patch location 2 simulations in the entire RVOT and only the transannular patch. (\textbf{B}) Enclosed volume by the stent across simulated conditions, separated by the distal region of the stent, the middle region of the stent, and the proximal region of the stent.}
    \label{fig:Patch_maximum_stress_strain}
\end{figure}

Figure \ref{fig:Patch_maximum_stress_strain}A highlights the maximum 1st principal stress and Lagrangian strain for all patch simulation cases in the entire RVOT and just the transannular patch. For patch location 1, the maximum stress was observed in the patched region. For patch location 2, maximum stress in the RVOT was observed in a location different from the patched region. Maximum strain in the entire RVOT was observed away from the patched region and was similar across all simulated conditions. Across both patched locations, maximum stress in the transannular patch increased, whereas maximum strain decreased as patch stiffness increased. Maximum stress and strain for the same patch stiffness values were greater in patch location 1 than in location 2. This location dependence is partly attributable to the underlying vessel geometry, and partly to the location of the patch relative to the stent. The patch contacts the stent frame in both locations, but the region of overlap differs. Location 1 lies within a tapered region of the RVOT, and the luminal layer of the patch coincides with the complete tapered distal end and middle region of the stent frame (see Figure \ref{fig:Patch_maximum_stress_strain}B). Location 2 lies within a flatter region of the RVOT, and the proximal half of the patch overlaps the proximal end of the stent, which flares to a wider diameter at its base (see Figure \ref{fig:Patch_maximum_stress_strain}B). Both features concentrate stress independently of patch stiffness, such that the elevated stress at location 1 reflects the combination of local geometry and stent-wall contact. The 95th\%ile, 75th\%ile, and mean first principal stresses and Lagrangian strains followed the same trends as maximum stress and Lagrangian strain.

The volume enclosed by the stent was calculated by dividing the stent into three regions: the distal end, the middle, and the proximal end, and is shown in Figure \ref{fig:Patch_maximum_stress_strain}B. Across all simulated conditions, the variation in the enclosed volume across these three regions was minimal.

\section{DISCUSSION}

We conducted a sensitivity analysis of the effect of RVOT material properties and material property spatial heterogeneity on the simulation of a self-expanding TPV deployment in the RVOT of a patient with ToF as a foundational step toward using simulations of TPVR to aid the assessment of patient candidacy and optimal device selection in patients with TPVR \cite{gillespie_patient_2017, mcelhinney_transcatheter_2024}. We found that vessel material properties significantly affect the resulting simulation, such as local stress and strain, but have a relatively lesser effect on overall vessel geometry. Spatial tissue property heterogeneity (such as inclusion of a stiff patch) had less effect on the resulting device conformation but significant effects on local vessel stress.

Simulation of self-expanding transcatheter valves has evolved over the last two decades and is now beginning to be clinically applied to adult TAVR. However, the first transcatheter valve deployed clinically was developed to address pulmonary regurgitation in patients with dysfunctional conduits in congenital heart disease, such as ToF \cite{bonhoeffer_transcatheter_2000}. Self-expanding devices such as the Harmony and Alterra pre-stent system now potentially meet the needs of the largest population with pulmonary insufficiency, represented by patients who have native outflow tracts after transannular patch placement in infancy.  In addition, multiple other devices and sizes of devices are in development \cite{jin_five-year_2024}. However, the new challenge is selecting which of these devices is optimal for an individual patient \cite{gillespie_patient_2017, mcelhinney_transcatheter_2024}.

Tissue mechanical properties can greatly influence the accuracy of simulations of self-expanding TPVR. Unfortunately, to date, experimentally-derived tissue material data for arteries in biaxial tensile testing has been limited to a small sample of patients, either infants \cite{cabrera_mechanical_2013} or older adults \cite{holzapfel_determination_2005}. Due to difficulty in the characterization of in vivo mechanical properties and the variability of the RVOT among different patients, it is important to understand the sensitivity of the constitutive material model parameters on the mechanical outcome metrics, such as stress and strain due to stent implantation, which may ultimately affect the clinical viability of the stent in a particular patient. To quantify the uncertainty in the simulation outcome, we performed a robust PCE-based uncertainty quantification analysis using UncertainSCI. Our sensitivity analysis revealed that the 95th\%ile stress was sensitive to the shear modulus of the ground matrix parameter $c$, the fiber modulus $k_1$, the fiber dispersion $\kappa$, and the fiber mean orientation angle $\gamma$, whereas 75th\%ile and mean stress were sensitive to $c$ and $\gamma$. 95th\%ile, 75th\%ile, and mean strain were only sensitive to $c$. Additionally, the first-order sensitivity indices had the greatest effect on total sensitivity indices, indicating that while important, secondary and higher-order indices do not have a significant enough effect on outcome metrics \cite{zhang_sobol_2015}. Despite variability in material properties, the greatest stress and strain regions were consistently at the locations where the proximal and distal ends of the TPV device impinge on the RVOT wall. Vessel material properties influenced the stress and strain values, but their effect on stress and strain distribution across the entire vessel and the expanded vessel geometry was minimal. 

Additional material variability that can be introduced in the RVOT wall includes the presence of prosthetic material introduced at the time of transannular patch during the initial surgical operation in infancy. On analyzing the effect of stent deployment in an RVOT with varying patch material stiffness, we found that the regions of greatest stress and least strain were at the junction where the patch embeds in the RVOT wall, irrespective of stiffness. Additionally, increasing the stiffness of the patch increased maximum stress and decreased maximum strain in the patch. These effects, however, should be interpreted in their geometric context as well. Higher stresses were observed in patch location 1, which lies within a tapered segment of the RVOT, contacting the distal end and middle region of the stent frame. Both features contributed to the maximum stress being localized to the transannular patch. In patch location 2, which is located in a flatter region of the RVOT, the maximum stress occurred away from the patch in a different region of the RVOT. Increasing patch stiffness, therefore, modulated the magnitude of these stress concentrations.

Importantly, these local stress differences did not translate into any appreciable change in device behavior. The volume enclosed by the stent was unaffected by the presence, location, or stiffness of the patch, and the response of the vessel within the stent landing zone varied negligibly across simulated conditions. While a stiff, heterogeneously located patch measurably alters the local stress distribution in the vessel wall, it does not change the deployed configuration of the device or the geometry of the landing zone in this model. This may suggest that the patch stiffness and location are more likely to bear on long-term, tissue-level considerations such as maladaptive remodeling at the patch-vessel junction than the immediate functional outcome of the deployment. These observations, nonetheless, support a role for simulation-based preprocedural assessment in anticipating where stress concentrations may arise.

There are a few limitations to this study. Firstly, the analyses were performed on a single patient-specific RVOT selected as a representative example. Native RVOT morphology is highly heterogeneous. Prior work has classified RVOTs into several distinct shape categories, which have been shown to influence device anchoring, oversizing, and procedural outcome \cite{schievano_variations_2007}. Extending the present workflow to a cohort spanning the recognized RVOT shape categories is an important next step, and multi-anatomy FE pipelines demonstrated for pulmonary conduits and RVOT pre-stenting provide a practical route to achieve this \cite{zelonis_integrated_2025, astori_fe-based_2025}.

Secondly, data on HGO material constants used in this study, which have been determined through either uniaxial \cite{jia_experimental_2017} or biaxial tensile testing \cite{cabrera_mechanical_2013} of pulmonary arteries, are limited to a small sample of patients in a wide age range. The samples of pulmonary arteries tested in these studies correspond to infants aged 8 months or adults aged 44-47 years. TPVR is generally performed for patients across a wide age range, which includes children, adolescents, young adults, and older adults \cite{nordmeyer_acute_2019, armstrong_association_2019, houeijeh_long-term_2023}. Therefore, there is a clinical need to evaluate tissue material properties for these age ranges. Constrained by the availability of appropriate tissue samples, future studies on determining elastic properties using 4D-CT-derived tissue deformation data can be a potential avenue to improve the fidelity of simulation results \cite{wu_noninvasive_2025}. Thirdly, the simulations conducted in this study evaluate biomechanical outcomes across conditions and do not consider blood flow or hemodynamics. Future work will leverage fluid-structure interaction to examine hemodynamic changes of TPV deployment, as has previously been conducted for TAVI \cite{luraghi_modeling_2019, ghosh_numerical_2020, basri_fluid_2020, fumagalli_fluidstructure_2023}. Fourthly, the RVOT model and simulations in this study only considered the diastolic phase of the cardiac cycle. Future work will incorporate both the systolic and diastolic phases to provide a comprehensive assessment over an entire cardiac cycle. Finally, the simulated conditions need to be validated with clinical TPVR procedure geometries. Future work will aim to include this validation to improve predicted outcomes.

\section{CONCLUSION} 
FE simulation provides a promising approach for evaluating optimal TPV placements before implantation and improving the success of repair. However, patient-specific tissue properties of the RVOT, which are critical for determining simulation outcomes, are often unknown. To better understand this limitation, we performed a sensitivity study to investigate the effects of material parameters on stress and strain in the RVOT model during TPV deployment. We found that RVOT stress and strain were most sensitive to uncertainties in the shear modulus of the ground matrix. While stress and strain values varied, their distribution across the vessel was similar. Furthermore, the vessel geometry on complete stent expansion was minimally affected by the material parameters. 

We further investigated model sensitivity for stress, strain, and RVOT enclosed volume in models with a transannular patch. Our results indicate that patch location and stiffness influenced both stress and strain values in the RVOT, while changes in enclosed volume by the stent were negligible. Because these findings were derived from a single representative RVOT anatomy, they should be confirmed across the range of native RVOT morphologies before being generalized. These collective findings suggest that although the location and material properties of the transannular patch may confound analysis of maladaptive tissue remodeling, FE simulation remains a reliable framework for evaluating immediate functional outcomes in TPVR.

\section{FUNDING}
This work was supported by NIH R01HL153166, K25HL168235, 2R01GM083925, the Additional Ventures Single Ventricle Research Fund, a CHOP Cardiac Center Innovation Award, the Topolewski Pediatric Valve Center at CHOP, and a CHOP Research Institute Post-Frontier Award.

\section{DISCLOSURES}
Matthew J. Gillespie is a consultant for Medtronic. The other authors have no disclosures.

\bibliographystyle{unsrt}
\bibliography{references_sensitivity.bib}

@article{schievano_variations_2007,
	title = {Variations in right ventricular outflow tract morphology following repair of congenital heart disease: implications for percutaneous pulmonary valve implantation},
	volume = {9},
	issn = {1097-6647},
	shorttitle = {Variations in right ventricular outflow tract morphology following repair of congenital heart disease},
	doi = {10.1080/10976640601187596},
	language = {eng},
	number = {4},
	journal = {Journal of Cardiovascular Magnetic Resonance: Official Journal of the Society for Cardiovascular Magnetic Resonance},
	author = {Schievano, Silvia and Coats, Louise and Migliavacca, Francesco and Norman, Wendy and Frigiola, Alessandra and Deanfield, John and Bonhoeffer, Philipp and Taylor, Andrew M.},
	year = {2007},
	pages = {687--695},
}

@article{benson_three-year_2020,
	title = {Three-{Year} {Outcomes} {From} the {Harmony} {Native} {Outflow} {Tract} {Early} {Feasibility} {Study}},
	volume = {13},
	issn = {1941-7632},
	doi = {10.1161/CIRCINTERVENTIONS.119.008320},
	language = {eng},
	number = {1},
	journal = {Circulation. Cardiovascular Interventions},
	author = {Benson, Lee N. and Gillespie, Matthew J. and Bergersen, Lisa and Cheatham, Sharon L. and Hor, Kan N. and Horlick, Eric M. and Weng, Shicheng and McHenry, Brian T. and Osten, Mark D. and Powell, Andrew J. and Cheatham, John P.},
	month = jan,
	year = {2020},
	pages = {e008320},
}

@article{zahn_first_2018,
	title = {First human implant of the {Alterra} {Adaptive} {PrestentTM} : {A} new self-expanding device designed to remodel the right ventricular outflow tract},
	volume = {91},
	issn = {1522-726X},
	shorttitle = {First human implant of the {Alterra} {Adaptive} {PrestentTM}},
	doi = {10.1002/ccd.27581},
	language = {eng},
	number = {6},
	journal = {Catheterization and Cardiovascular Interventions: Official Journal of the Society for Cardiac Angiography \& Interventions},
	author = {Zahn, Evan M. and Chang, Jennifer C. and Armer, Dustin and Garg, Ruchira},
	month = may,
	year = {2018},
	pages = {1125--1129},
}

@article{jin_five-year_2024,
	title = {Five-year follow-up after percutaneous pulmonary valve implantation using the {Venus} {P}-valve system for patients with pulmonary regurgitation and an enlarged native right ventricular outflow tract},
	volume = {103},
	issn = {1522-726X},
	doi = {10.1002/ccd.30916},
	language = {eng},
	number = {2},
	journal = {Catheterization and Cardiovascular Interventions: Official Journal of the Society for Cardiac Angiography \& Interventions},
	author = {Jin, Qinchun and Long, Yuliang and Zhang, Gejun and Pan, Xin and Chen, Mao and Feng, Yuan and Liu, Jinfen and Yu, Shiqiang and Pan, Wenzhi and Zhou, Daxin and Ge, Junbo},
	month = feb,
	year = {2024},
	pages = {359--366},
}

@article{gillespie_patient_2017,
	title = {Patient {Selection} {Process} for the {Harmony} {Transcatheter} {Pulmonary} {Valve} {Early} {Feasibility} {Study}},
	volume = {120},
	issn = {1879-1913},
	doi = {10.1016/j.amjcard.2017.07.034},
	language = {eng},
	number = {8},
	journal = {The American Journal of Cardiology},
	author = {Gillespie, Matthew J. and Benson, Lee N. and Bergersen, Lisa and Bacha, Emile A. and Cheatham, Sharon L. and Crean, Andrew M. and Eicken, Andreas and Ewert, Peter and Geva, Tal and Hellenbrand, William E. and Hor, Kan N. and Horlick, Eric M. and Jones, Thomas K. and Mayer, John and McHenry, Brian T. and Osten, Mark D. and Powell, Andrew J. and Zahn, Evan M. and Cheatham, John P.},
	month = oct,
	year = {2017},
	pages = {1387--1392},
}

@article{mcelhinney_transcatheter_2024,
	title = {Transcatheter {Pulmonary} {Valve} {Replacement} {With} the {Harmony} {Valve} in {Patients} {Who} {Do} {Not} {Meet} {Recommended} {Oversizing} {Criteria} on the {Screening} {Perimeter} {Plot}},
	volume = {17},
	issn = {1941-7632},
	doi = {10.1161/CIRCINTERVENTIONS.123.013889},
	language = {eng},
	number = {5},
	journal = {Circulation. Cardiovascular Interventions},
	author = {McElhinney, Doff B. and Gillespie, Matthew J. and Aboulhosn, Jamil A. and Cabalka, Allison K. and Morray, Brian H. and Balzer, David T. and Qureshi, Athar M. and Hoskoppal, Arvind K. and Goldstein, Bryan H.},
	month = may,
	year = {2024},
	pages = {e013889},
}

@article{gillespie_1-year_2023,
	title = {1-{Year} {Outcomes} in a {Pooled} {Cohort} of {Harmony} {Transcatheter} {Pulmonary} {Valve} {Clinical} {Trial} {Participants}},
	volume = {16},
	issn = {19368798},
	url = {https://linkinghub.elsevier.com/retrieve/pii/S1936879823005617},
	doi = {10.1016/j.jcin.2023.03.002},
	language = {en},
	number = {15},
	urldate = {2025-06-06},
	journal = {JACC: Cardiovascular Interventions},
	author = {Gillespie, Matthew J. and McElhinney, Doff B. and Jones, Thomas K. and Levi, Daniel S. and Asnes, Jeremy and Gray, Robert G. and Cabalka, Allison K. and Fujimoto, Kazuto and Qureshi, Athar M. and Justino, Henri and Bergersen, Lisa and Benson, Lee N. and Haugan, Daniel and Boe, Brian A. and Cheatham, John P.},
	month = aug,
	year = {2023},
	pages = {1917--1928},
}

@article{gartenberg_transcatheter_2023,
	title = {Transcatheter {Approaches} to {Pulmonary} {Valve} {Replacement} in {Congenital} {Heart} {Disease}: {Revolutionizing} the {Management} of {RVOT} {Dysfunction}?},
	volume = {35},
	issn = {10430679},
	shorttitle = {Transcatheter {Approaches} to {Pulmonary} {Valve} {Replacement} in {Congenital} {Heart} {Disease}},
	url = {https://linkinghub.elsevier.com/retrieve/pii/S104306792200048X},
	doi = {10.1053/j.semtcvs.2022.02.009},
	language = {en},
	number = {2},
	urldate = {2025-06-06},
	journal = {Seminars in Thoracic and Cardiovascular Surgery},
	author = {Gartenberg, Ari J. and Gillespie, Matthew J. and Glatz, Andrew C.},
	year = {2023},
	pages = {333--338},
}

@article{kenny_current_2017,
	title = {Current {Status} and {Future} {Potential} of {Transcatheter} {Interventions} in {Congenital} {Heart} {Disease}},
	volume = {120},
	issn = {0009-7330, 1524-4571},
	url = {https://www.ahajournals.org/doi/10.1161/CIRCRESAHA.116.309185},
	doi = {10.1161/CIRCRESAHA.116.309185},
	language = {en},
	number = {6},
	urldate = {2025-06-06},
	journal = {Circulation Research},
	author = {Kenny, Damien P. and Hijazi, Ziyad M.},
	month = mar,
	year = {2017},
	pages = {1015--1026},
}

@article{patel_transcatheter_2022,
	title = {Transcatheter {Pulmonary} {Valve} {Replacement}: {A} {Review} of {Current} {Valve} {Technologies}},
	volume = {1},
	issn = {27729303},
	shorttitle = {Transcatheter {Pulmonary} {Valve} {Replacement}},
	url = {https://linkinghub.elsevier.com/retrieve/pii/S2772930322004501},
	doi = {10.1016/j.jscai.2022.100452},
	language = {en},
	number = {6},
	urldate = {2025-06-06},
	journal = {Journal of the Society for Cardiovascular Angiography \& Interventions},
	author = {Patel, Neil D. and Levi, Daniel S. and Cheatham, John P. and Qureshi, Shakeel A. and Shahanavaz, Shabana and Zahn, Evan M.},
	month = nov,
	year = {2022},
	pages = {100452},
}

@article{capelli_patient-specific_2010,
	title = {Patient-specific reconstructed anatomies and computer simulations are fundamental for selecting medical device treatment: application to a new percutaneous pulmonary valve},
	volume = {368},
	issn = {1364-503X, 1471-2962},
	shorttitle = {Patient-specific reconstructed anatomies and computer simulations are fundamental for selecting medical device treatment},
	url = {https://royalsocietypublishing.org/doi/10.1098/rsta.2010.0088},
	doi = {10.1098/rsta.2010.0088},
	language = {en},
	number = {1921},
	urldate = {2025-06-06},
	journal = {Philosophical Transactions of the Royal Society A: Mathematical, Physical and Engineering Sciences},
	author = {Capelli, Claudio and Taylor, Andrew M. and Migliavacca, Francesco and Bonhoeffer, Philipp and Schievano, Silvia},
	month = jun,
	year = {2010},
	pages = {3027--3038},
}

@misc{medtronic_tpv_2021,
	title = {{TPV} 22 and {TPV} 25 {Transcatheter} {Pulmonary} {Valves} and {Delivery} {Catheter} {System}},
	url = {https://www.accessdata.fda.gov/cdrh_docs/pdf20/P200046D.pdf},
	author = {Medtronic},
	year = {2021},
}

@article{donahue_finite_2024,
	title = {Finite element modeling with patient‐specific geometry to assess clinical risks of percutaneous pulmonary valve implantation},
	volume = {103},
	issn = {1522-1946, 1522-726X},
	url = {https://onlinelibrary.wiley.com/doi/10.1002/ccd.31016},
	doi = {10.1002/ccd.31016},
	language = {en},
	number = {6},
	urldate = {2025-06-06},
	journal = {Catheterization and Cardiovascular Interventions},
	author = {Donahue, Carly L. and Westman, Claire L. and Faanes, Brittany L. and Qureshi, Athar M. and Barocas, Victor H. and Aggarwal, Varun},
	month = may,
	year = {2024},
	pages = {924--933},
}

@article{vanderveken_mechano-biological_2020,
	title = {Mechano-biological adaptation of the pulmonary artery exposed to systemic conditions},
	volume = {10},
	issn = {2045-2322},
	url = {https://www.nature.com/articles/s41598-020-59554-7},
	doi = {10.1038/s41598-020-59554-7},
	language = {en},
	number = {1},
	urldate = {2025-06-06},
	journal = {Scientific Reports},
	author = {Vanderveken, Emma and Vastmans, Julie and Claus, Piet and Verbeken, Eric and Fehervary, Heleen and Van Hoof, Lucas and Vandendriessche, Katrien and Verbrugghe, Peter and Famaey, Nele and Rega, Filip},
	month = feb,
	year = {2020},
	pages = {2724},
}

@article{holzapfel_new_2000,
	title = {A {New} {Constitutive} {Framework} for {Arterial} {Wall} {Mechanics} and a {Comparative} {Study} of {Material} {Models}},
	volume = {61},
	issn = {03743535},
	url = {http://link.springer.com/10.1023/A:1010835316564},
	doi = {10.1023/A:1010835316564},
	number = {1/3},
	urldate = {2025-06-06},
	journal = {Journal of Elasticity},
	author = {Holzapfel, Gerhard A. and Gasser, Thomas C. and Ogden, Ray W.},
	year = {2000},
	pages = {1--48},
}

@article{gasser_hyperelastic_2006,
	title = {Hyperelastic modelling of arterial layers with distributed collagen fibre orientations},
	volume = {3},
	copyright = {https://royalsociety.org/journals/ethics-policies/data-sharing-mining/},
	issn = {1742-5689, 1742-5662},
	url = {https://royalsocietypublishing.org/doi/10.1098/rsif.2005.0073},
	doi = {10.1098/rsif.2005.0073},
	language = {en},
	number = {6},
	urldate = {2025-06-06},
	journal = {Journal of The Royal Society Interface},
	author = {Gasser, T. Christian and Ogden, Ray W and Holzapfel, Gerhard A},
	month = feb,
	year = {2006},
	pages = {15--35},
}

@inproceedings{donahue_finite_2022,
	address = {Minneapolis, MN, USA},
	title = {Finite {Element} {Modeling} {Using} {Patient}-{Specific} {Geometry} to {Predict} {Aortic} {Valve} {Insufficiency} {During} {Percutaneous} {Pulmonary} {Valve} {Implantation}},
	isbn = {978-0-7918-8571-0},
	url = {https://asmedigitalcollection.asme.org/BIOMED/proceedings/DMD2022/84815/V001T02A001/1140616},
	doi = {10.1115/DMD2022-1022},
	urldate = {2025-06-06},
	booktitle = {2022 {Design} of {Medical} {Devices} {Conference}},
	publisher = {American Society of Mechanical Engineers},
	author = {Donahue, Carly L. and Aggarwal, Varun and Barocas, Victor H.},
	month = apr,
	year = {2022},
	pages = {V001T02A001},
}

@article{vemury_behaviour_2019,
	title = {The behaviour of {Nitinol} {Wire} {Bundles} for {Structural} {Applications}},
	volume = {03},
	issn = {26895846},
	url = {http://www.lidsen.com/journals/rpm/rpm-03-01-009},
	doi = {10.21926/rpm.2101009},
	number = {01},
	urldate = {2025-06-06},
	journal = {Recent Progress in Materials},
	author = {Vemury, Chandra Mouli and Corradi, Marco and Abozaid, Feras and Charles, Alasdair and Hughes, David},
	month = aug,
	year = {2019},
	pages = {1--1},
}

@article{holzapfel_determination_2005,
	title = {Determination of layer-specific mechanical properties of human coronary arteries with nonatherosclerotic intimal thickening and related constitutive modeling},
	volume = {289},
	issn = {0363-6135},
	doi = {10.1152/ajpheart.00934.2004},
	language = {eng},
	number = {5},
	journal = {American Journal of Physiology. Heart and Circulatory Physiology},
	author = {Holzapfel, Gerhard A. and Sommer, Gerhard and Gasser, Christian T. and Regitnig, Peter},
	month = nov,
	year = {2005},
	pages = {H2048--2058},
}

@article{diaz-pinto_monai_2024,
	title = {{MONAI} {Label}: {A} framework for {AI}-assisted interactive labeling of {3D} medical images},
	volume = {95},
	issn = {13618415},
	shorttitle = {{MONAI} {Label}},
	url = {https://linkinghub.elsevier.com/retrieve/pii/S1361841524001324},
	doi = {10.1016/j.media.2024.103207},
	language = {en},
	urldate = {2025-06-07},
	journal = {Medical Image Analysis},
	author = {Diaz-Pinto, Andres and Alle, Sachidanand and Nath, Vishwesh and Tang, Yucheng and Ihsani, Alvin and Asad, Muhammad and Pérez-García, Fernando and Mehta, Pritesh and Li, Wenqi and Flores, Mona and Roth, Holger R. and Vercauteren, Tom and Xu, Daguang and Dogra, Prerna and Ourselin, Sebastien and Feng, Andrew and Cardoso, M. Jorge},
	month = jul,
	year = {2024},
	pages = {103207},
}

@misc{maas_uncoupled_2025,
	title = {Uncoupled {Holzapfel}-{Gasser}-{Ogden}},
	url = {https://help.febio.org/docs/FEBioUser-4-9/UM49-4.1.2.8.html},
	publisher = {FEBio User's Manual 4.9},
	author = {Maas, Steve A. and Ateshian, Gerard A. and Weiss, Jeffrey A. and Herron, Michael},
	month = mar,
	year = {2025},
}

@misc{maas_isotropic_2025,
	title = {Isotropic {Elastic}},
	url = {https://help.febio.org/docs/FEBioUser-4-9/UM49-4.1.4.14.html},
	publisher = {FEBio User's Manual 4.9},
	author = {Maas, Steve A. and Ateshian, Gerard A. and Weiss, Jeffrey A. and Herron, Michael},
	month = mar,
	year = {2025},
}

@misc{maas_neo-hookean_2025,
	title = {Neo-{Hookean}},
	url = {https://help.febio.org/docs/FEBioUser-4-9/UM49-4.1.4.19.html},
	publisher = {FEBio User's Manual 4.9},
	author = {Maas, Steve A. and Ateshian, Gerard A. and Weiss, Jeffrey A. and Herron, Michael},
	month = mar,
	year = {2025},
}

@article{schoonbeek_implantation_2016,
	title = {Implantation of the {Medtronic} {Harmony} {Transcatheter} {Pulmonary} {Valve} {Improves} {Right} {Ventricular} {Size} and {Function} in an {Ovine} {Model} of {Postoperative} {Chronic} {Pulmonary} {Insufficiency}},
	volume = {9},
	issn = {1941-7640, 1941-7632},
	url = {https://www.ahajournals.org/doi/10.1161/CIRCINTERVENTIONS.116.003920},
	doi = {10.1161/circinterventions.116.003920},
	language = {en},
	number = {10},
	urldate = {2025-07-17},
	journal = {Circulation: Cardiovascular Interventions},
	publisher = {Ovid Technologies (Wolters Kluwer Health)},
	author = {Schoonbeek, Rosanne C. and Takebayashi, Satoshi and Aoki, Chikashi and Shimaoka, Toru and Harris, Matthew A. and Fu, Gregory L. and Kim, Timothy S. and Dori, Yoav and McGarvey, Jeremy and Litt, Harold and Bouma, Wobbe and Zsido, Gerald and Glatz, Andrew C. and Rome, Jonathan J. and Gorman, Robert C. and Gorman, Joseph H. and Gillespie, Matthew J.},
	month = oct,
	year = {2016},
}

@article{biernacka_transcatheter_2015,
	title = {Transcatheter pulmonary valve implantation in patients with right ventricular outflow tract dysfunction: early and mid-term results},
	volume = {27},
	issn = {1557-2501},
	shorttitle = {Transcatheter pulmonary valve implantation in patients with right ventricular outflow tract dysfunction},
	language = {eng},
	number = {6},
	journal = {The Journal of Invasive Cardiology},
	author = {Biernacka, Elżbieta Katarzyna and Rużyłło, Witold and Demkow, Marcin and Kowalski, Mirosław and Śpiewak, Mateusz and Piotrowski, Walerian and Kuśmierczyk, Mariusz and Banaś, Sławomir and Różanski, Jacek and Hoffman, Piotr},
	month = jun,
	year = {2015},
	pages = {E82--89},
}

@article{demkow_percutaneous_2014,
	title = {Percutaneous edwards {SAPIEN}$^{\textrm{™}}$ valve implantation for significant pulmonary regurgitation after previous surgical repair with a right ventricular outflow patch},
	volume = {83},
	copyright = {http://onlinelibrary.wiley.com/termsAndConditions\#vor},
	issn = {1522-1946, 1522-726X},
	url = {https://onlinelibrary.wiley.com/doi/10.1002/ccd.25096},
	doi = {10.1002/ccd.25096},
	language = {en},
	number = {3},
	urldate = {2025-07-17},
	journal = {Catheterization and Cardiovascular Interventions},
	publisher = {Wiley},
	author = {Demkow, Marcin and Rużyłło, Witold and Biernacka, Elżbieta Katarzyna and Kalińczuk, Łukasz and Śpiewak, Mateusz and Kowalski, Mirosław and Sitkowska, Ewa and Kuśmierczyk, Mariusz and Różanski, Jacek and Banaś, Sławomir and Chmielak, Zbigniew and Hoffman, Piotr},
	month = feb,
	year = {2014},
	pages = {474--481},
}

@article{ghawi_transcatheter_2012,
	title = {Transcatheter {Pulmonary} {Valve} {Replacement}},
	volume = {1},
	issn = {2193-8261, 2193-6544},
	url = {http://link.springer.com/10.1007/s40119-012-0005-9},
	doi = {10.1007/s40119-012-0005-9},
	language = {en},
	number = {1},
	urldate = {2025-07-17},
	journal = {Cardiology and Therapy},
	publisher = {Springer Science and Business Media LLC},
	author = {Ghawi, Hani and Kenny, Damien and Hijazi, Ziyad M.},
	month = dec,
	year = {2012},
}

@article{kenny_percutaneous_2011,
	title = {Percutaneous {Implantation} of the {Edwards} {SAPIEN} {Transcatheter} {Heart} {Valve} for {Conduit} {Failure} in the {Pulmonary} {Position}},
	volume = {58},
	copyright = {https://www.elsevier.com/tdm/userlicense/1.0/},
	issn = {0735-1097},
	url = {https://linkinghub.elsevier.com/retrieve/pii/S0735109711030828},
	doi = {10.1016/j.jacc.2011.07.040},
	language = {en},
	number = {21},
	urldate = {2025-07-17},
	journal = {Journal of the American College of Cardiology},
	publisher = {Elsevier BV},
	author = {Kenny, Damien and Hijazi, Ziyad M. and Kar, Saibal and Rhodes, John and Mullen, Michael and Makkar, Raj and Shirali, Girish and Fogel, Mark and Fahey, John and Heitschmidt, Mary G. and Cain, Christopher},
	month = nov,
	year = {2011},
	pages = {2248--2256},
}

@article{obyrne_trends_2015,
	title = {Trends in {Pulmonary} {Valve} {Replacement} in {Children} and {Adults} {With} {Tetralogy} of {Fallot}},
	volume = {115},
	copyright = {https://www.elsevier.com/tdm/userlicense/1.0/},
	issn = {0002-9149},
	url = {https://linkinghub.elsevier.com/retrieve/pii/S0002914914019389},
	doi = {10.1016/j.amjcard.2014.09.054},
	language = {en},
	number = {1},
	urldate = {2025-07-17},
	journal = {The American Journal of Cardiology},
	publisher = {Elsevier BV},
	author = {O’Byrne, Michael L. and Glatz, Andrew C. and Mercer-Rosa, Laura and Gillespie, Matthew J. and Dori, Yoav and Goldmuntz, Elizabeth and Kawut, Steven and Rome, Jonathan J.},
	month = jan,
	year = {2015},
	pages = {118--124},
}

@article{valente_contemporary_2014,
	title = {Contemporary predictors of death and sustained ventricular tachycardia in patients with repaired tetralogy of {Fallot} enrolled in the {INDICATOR} cohort},
	volume = {100},
	issn = {1355-6037, 1468-201X},
	url = {https://heart.bmj.com/lookup/doi/10.1136/heartjnl-2013-304958},
	doi = {10.1136/heartjnl-2013-304958},
	language = {en},
	number = {3},
	urldate = {2025-07-17},
	journal = {Heart},
	publisher = {BMJ},
	author = {Valente, Anne Marie and Gauvreau, Kimberlee and Assenza, Gabriele Egidy and Babu-Narayan, Sonya V and Schreier, Jenna and Gatzoulis, Michael A and Groenink, Maarten and Inuzuka, Ryo and Kilner, Philip J and Koyak, Zeliha and Landzberg, Michael J and Mulder, Barbara and Powell, Andrew J and Wald, Rachel and Geva, Tal},
	month = feb,
	year = {2014},
	pages = {247--253},
}

@article{lasso_slicerheart_2022,
	title = {{SlicerHeart}: {An} open-source computing platform for cardiac image analysis and modeling},
	volume = {9},
	copyright = {https://creativecommons.org/licenses/by/4.0/},
	issn = {2297-055X},
	shorttitle = {{SlicerHeart}},
	url = {https://www.frontiersin.org/articles/10.3389/fcvm.2022.886549/full},
	doi = {10.3389/fcvm.2022.886549},
	urldate = {2025-07-17},
	journal = {Frontiers in Cardiovascular Medicine},
	publisher = {Frontiers Media SA},
	author = {Lasso, Andras and Herz, Christian and Nam, Hannah and Cianciulli, Alana and Pieper, Steve and Drouin, Simon and Pinter, Csaba and St-Onge, Samuelle and Vigil, Chad and Ching, Stephen and Sunderland, Kyle and Fichtinger, Gabor and Kikinis, Ron and Jolley, Matthew A.},
	month = sep,
	year = {2022},
}

@article{maas_febio_2012,
	title = {{FEBio}: {Finite} {Elements} for {Biomechanics}},
	volume = {134},
	issn = {0148-0731, 1528-8951},
	shorttitle = {{FEBio}},
	url = {https://asmedigitalcollection.asme.org/biomechanical/article/doi/10.1115/1.4005694/455684/FEBio-Finite-Elements-for-Biomechanics},
	doi = {10.1115/1.4005694},
	language = {en},
	number = {1},
	urldate = {2025-07-17},
	journal = {Journal of Biomechanical Engineering},
	publisher = {ASME International},
	author = {Maas, Steve A. and Ellis, Benjamin J. and Ateshian, Gerard A. and Weiss, Jeffrey A.},
	month = jan,
	year = {2012},
}

@article{burk_efficient_2020,
	title = {Efficient sampling for polynomial chaos‐based uncertainty quantification and sensitivity analysis using weighted approximate {Fekete} points},
	volume = {36},
	copyright = {http://onlinelibrary.wiley.com/termsAndConditions\#vor},
	issn = {2040-7939, 2040-7947},
	url = {https://onlinelibrary.wiley.com/doi/10.1002/cnm.3395},
	doi = {10.1002/cnm.3395},
	language = {en},
	number = {11},
	urldate = {2025-07-17},
	journal = {International Journal for Numerical Methods in Biomedical Engineering},
	publisher = {Wiley},
	author = {Burk, Kyle M. and Narayan, Akil and Orr, Joseph A.},
	month = nov,
	year = {2020},
}

@article{jolley_toward_2019,
	title = {Toward predictive modeling of catheter‐based pulmonary valve replacement into native right ventricular outflow tracts},
	volume = {93},
	copyright = {http://onlinelibrary.wiley.com/termsAndConditions\#vor},
	issn = {1522-1946, 1522-726X},
	url = {https://onlinelibrary.wiley.com/doi/10.1002/ccd.27962},
	doi = {10.1002/ccd.27962},
	language = {en},
	number = {3},
	urldate = {2025-07-17},
	journal = {Catheterization and Cardiovascular Interventions},
	publisher = {Wiley},
	author = {Jolley, Matthew A. and Lasso, Andras and Nam, Hannah H. and Dinh, Patrick V. and Scanlan, Adam B. and Nguyen, Alex V. and Ilina, Anna and Morray, Brian and Glatz, Andrew C. and McGowan, Francis X. and Whitehead, Kevin and Dori, Yoav and Gorman, Joseph H. and Gorman, Robert C. and Fichtinger, Gabor and Gillespie, Matthew J.},
	month = feb,
	year = {2019},
}

@article{wu_effects_2023,
	title = {The effects of leaflet material properties on the simulated function of regurgitant mitral valves},
	volume = {142},
	copyright = {https://www.elsevier.com/tdm/userlicense/1.0/},
	issn = {1751-6161},
	url = {https://linkinghub.elsevier.com/retrieve/pii/S1751616123002114},
	doi = {10.1016/j.jmbbm.2023.105858},
	language = {en},
	urldate = {2025-07-17},
	journal = {Journal of the Mechanical Behavior of Biomedical Materials},
	publisher = {Elsevier BV},
	author = {Wu, Wensi and Ching, Stephen and Sabin, Patricia and Laurence, Devin W. and Maas, Steve A. and Lasso, Andras and Weiss, Jeffrey A. and Jolley, Matthew A.},
	month = jun,
	year = {2023},
	pages = {105858},
}

@article{wu_computational_2022,
	title = {A {Computational} {Framework} for {Atrioventricular} {Valve} {Modeling} {Using} {Open}-{Source} {Software}},
	volume = {144},
	copyright = {https://www.asme.org/publications-submissions/publishing-information/legal-policies},
	issn = {0148-0731, 1528-8951},
	url = {https://asmedigitalcollection.asme.org/biomechanical/article/144/10/101012/1140708/A-Computational-Framework-for-Atrioventricular},
	doi = {10.1115/1.4054485},
	language = {en},
	number = {10},
	urldate = {2025-07-17},
	journal = {Journal of Biomechanical Engineering},
	publisher = {ASME International},
	author = {Wu, Wensi and Ching, Stephen and Maas, Steve A. and Lasso, Andras and Sabin, Patricia and Weiss, Jeffrey A. and Jolley, Matthew A.},
	month = oct,
	year = {2022},
}

@article{volokh_modeling_2011,
	title = {Modeling failure of soft anisotropic materials with application to arteries},
	volume = {4},
	copyright = {https://www.elsevier.com/tdm/userlicense/1.0/},
	issn = {1751-6161},
	url = {https://linkinghub.elsevier.com/retrieve/pii/S1751616111000038},
	doi = {10.1016/j.jmbbm.2011.01.002},
	language = {en},
	number = {8},
	urldate = {2025-07-17},
	journal = {Journal of the Mechanical Behavior of Biomedical Materials},
	publisher = {Elsevier BV},
	author = {Volokh, K.Y.},
	month = nov,
	year = {2011},
	pages = {1582--1594},
}

@article{jia_experimental_2017,
	title = {Experimental {Study} of {Anisotropic} {Stress}/{Strain} {Relationships} of {Aortic} and {Pulmonary} {Artery} {Homografts} and {Synthetic} {Vascular} {Grafts}},
	volume = {139},
	issn = {0148-0731, 1528-8951},
	url = {https://asmedigitalcollection.asme.org/biomechanical/article/doi/10.1115/1.4037400/371335/Experimental-Study-of-Anisotropic-StressStrain},
	doi = {10.1115/1.4037400},
	language = {en},
	number = {10},
	urldate = {2025-07-17},
	journal = {Journal of Biomechanical Engineering},
	publisher = {ASME International},
	author = {Jia, Yueqian and Qiao, Yangyang and Ricardo Argueta-Morales, I. and Maung, Aung and Norfleet, Jack and Bai, Yuanli and Divo, Eduardo and Kassab, Alain J. and DeCampli, William M.},
	month = oct,
	year = {2017},
}

@article{cabrera_mechanical_2013,
	title = {Mechanical analysis of ovine and pediatric pulmonary artery for heart valve stent design},
	volume = {46},
	copyright = {https://www.elsevier.com/tdm/userlicense/1.0/},
	issn = {0021-9290},
	url = {https://linkinghub.elsevier.com/retrieve/pii/S0021929013002145},
	doi = {10.1016/j.jbiomech.2013.04.020},
	language = {en},
	number = {12},
	urldate = {2025-07-17},
	journal = {Journal of Biomechanics},
	publisher = {Elsevier BV},
	author = {Cabrera, M.S. and Oomens, C.W.J. and Bouten, C.V.C. and Bogers, A.J.J.C. and Hoerstrup, S.P. and Baaijens, F.P.T.},
	month = aug,
	year = {2013},
	pages = {2075--2081},
}

@article{wu_noninvasive_2025,
	title = {A noninvasive method for determining elastic parameters of valve tissue using physics-informed neural networks},
	volume = {200},
	copyright = {https://www.elsevier.com/tdm/userlicense/1.0/},
	issn = {1742-7061},
	url = {https://linkinghub.elsevier.com/retrieve/pii/S1742706125003472},
	doi = {10.1016/j.actbio.2025.05.021},
	language = {en},
	urldate = {2025-07-17},
	journal = {Acta Biomaterialia},
	publisher = {Elsevier BV},
	author = {Wu, Wensi and Daneker, Mitchell and Herz, Christian and Dewey, Hannah and Weiss, Jeffrey A. and Pouch, Alison M. and Lu, Lu and Jolley, Matthew A.},
	month = jun,
	year = {2025},
	pages = {283--298},
}

@article{rosenthal_aneurysms_1972,
	title = {Aneurysms of right ventricular outflow patches},
	volume = {63},
	copyright = {https://www.elsevier.com/tdm/userlicense/1.0/},
	issn = {0022-5223},
	url = {https://linkinghub.elsevier.com/retrieve/pii/S0022522319418412},
	doi = {10.1016/s0022-5223(19)41841-2},
	language = {en},
	number = {5},
	urldate = {2025-07-17},
	journal = {The Journal of Thoracic and Cardiovascular Surgery},
	publisher = {Elsevier BV},
	author = {Rosenthal, Amnon and Gross, Robert E. and Pasternac, Andre},
	month = may,
	year = {1972},
	pages = {735--740},
}

@article{mc_dynamic_2015,
	title = {Dynamic and fluid-structure interaction simulations of bioprosthetic heart valves using parametric design with {T}-splines and {Fung}-type material models},
	volume = {55},
	issn = {0178-7675},
	url = {https://pubmed.ncbi.nlm.nih.gov/26392645/},
	doi = {10.1007/s00466-015-1166-x},
	language = {en},
	number = {6},
	urldate = {2025-07-26},
	journal = {Computational mechanics},
	author = {Mc, Hsu and D, Kamensky and F, Xu and J, Kiendl and C, Wang and Mc, Wu and J, Mineroff and A, Reali and Y, Bazilevs and Ms, Sacks},
	month = jun,
	year = {2015},
}

@article{m_measurement_2023,
	title = {The {Measurement} of {Bovine} {Pericardium} {Density} and {Its} {Implications} on {Leaflet} {Stress} {Distribution} in {Bioprosthetic} {Heart} {Valves}},
	volume = {14},
	issn = {1869-4098},
	url = {https://pubmed.ncbi.nlm.nih.gov/37932655/},
	doi = {10.1007/s13239-023-00692-0},
	language = {en},
	number = {6},
	urldate = {2025-07-26},
	journal = {Cardiovascular engineering and technology},
	author = {M, Sadipour and An, Azadani},
	month = dec,
	year = {2023},
}

@misc{zelonis_integrated_2025,
	title = {Integrated {Open}-{Source} {Framework} for {Simulation} of {Transcatheter} {Pulmonary} {Valves} in {Native} {Right} {Ventricular} {Outflow} {Tracts}},
	copyright = {Creative Commons Attribution 4.0 International},
	url = {https://arxiv.org/abs/2507.06337},
	doi = {10.48550/ARXIV.2507.06337},
	urldate = {2025-08-21},
	publisher = {arXiv},
	author = {Zelonis, Christopher N. and Maheshwari, Jalaj and Wu, Wensi and Maas, Steve A. and Aslan, Seda and Sunderland, Kyle and Ching, Stephen and Koluda, Ashley and Barak-Corren, Yuval and Mangine, Nicolas and Sabin, Patricia M. and Lasso, Andras and Laurence, Devin W. and Herz, Christian and Gillespie, Matthew J. and Weiss, Jeffrey A. and Jolley, Matthew A.},
	year = {2025},
	note = {Version Number: 2},
}

@article{bonhoeffer_transcatheter_2000,
	title = {Transcatheter {Implantation} of a {Bovine} {Valve} in {Pulmonary} {Position}: {A} {Lamb} {Study}},
	volume = {102},
	issn = {0009-7322, 1524-4539},
	shorttitle = {Transcatheter {Implantation} of a {Bovine} {Valve} in {Pulmonary} {Position}},
	url = {https://www.ahajournals.org/doi/10.1161/01.CIR.102.7.813},
	doi = {10.1161/01.CIR.102.7.813},
	language = {en},
	number = {7},
	urldate = {2025-08-21},
	journal = {Circulation},
	author = {Bonhoeffer, Philipp and Boudjemline, Younes and Saliba, Zakhia and Hausse, Ana Olga and Aggoun, Yacine and Bonnet, Damien and Sidi, Daniel and Kachaner, Jean},
	month = aug,
	year = {2000},
	pages = {813--816},
}

@article{morganti_simulation_2014,
	title = {Simulation of transcatheter aortic valve implantation through patient-specific finite element analysis: {Two} clinical cases},
	volume = {47},
	issn = {00219290},
	shorttitle = {Simulation of transcatheter aortic valve implantation through patient-specific finite element analysis},
	url = {https://linkinghub.elsevier.com/retrieve/pii/S0021929014003467},
	doi = {10.1016/j.jbiomech.2014.06.007},
	language = {en},
	number = {11},
	urldate = {2025-08-28},
	journal = {Journal of Biomechanics},
	author = {Morganti, S. and Conti, M. and Aiello, M. and Valentini, A. and Mazzola, A. and Reali, A. and Auricchio, F.},
	month = aug,
	year = {2014},
	pages = {2547--2555},
}

@article{ovcharenko_modeling_2016,
	title = {Modeling of transcatheter aortic valve replacement: {Patient} specific vs general approaches based on finite element analysis},
	volume = {69},
	issn = {00104825},
	shorttitle = {Modeling of transcatheter aortic valve replacement},
	url = {https://linkinghub.elsevier.com/retrieve/pii/S0010482515003844},
	doi = {10.1016/j.compbiomed.2015.12.001},
	language = {en},
	urldate = {2025-08-28},
	journal = {Computers in Biology and Medicine},
	author = {Ovcharenko, E.A. and Klyshnikov, K.U. and Yuzhalin, A.E. and Savrasov, G.V. and Kokov, A.N. and Batranin, A.V. and Ganyukov, V.I. and Kudryavtseva, Y.A.},
	month = feb,
	year = {2016},
	pages = {29--36},
}

@article{auricchio_simulation_2014,
	title = {Simulation of transcatheter aortic valve implantation: a patient-specific finite element approach},
	volume = {17},
	issn = {1025-5842, 1476-8259},
	shorttitle = {Simulation of transcatheter aortic valve implantation},
	url = {http://www.tandfonline.com/doi/abs/10.1080/10255842.2012.746676},
	doi = {10.1080/10255842.2012.746676},
	language = {en},
	number = {12},
	urldate = {2025-08-28},
	journal = {Computer Methods in Biomechanics and Biomedical Engineering},
	author = {Auricchio, F. and Conti, M. and Morganti, S. and Reali, A.},
	month = sep,
	year = {2014},
	pages = {1347--1357},
}

@article{sturla_impact_2016,
	title = {Impact of different aortic valve calcification patterns on the outcome of transcatheter aortic valve implantation: {A} finite element study},
	volume = {49},
	issn = {00219290},
	shorttitle = {Impact of different aortic valve calcification patterns on the outcome of transcatheter aortic valve implantation},
	url = {https://linkinghub.elsevier.com/retrieve/pii/S0021929016303669},
	doi = {10.1016/j.jbiomech.2016.03.036},
	language = {en},
	number = {12},
	urldate = {2025-08-28},
	journal = {Journal of Biomechanics},
	author = {Sturla, Francesco and Ronzoni, Mattia and Vitali, Mattia and Dimasi, Annalisa and Vismara, Riccardo and Preston-Maher, Georgia and Burriesci, Gaetano and Votta, Emiliano and Redaelli, Alberto},
	month = aug,
	year = {2016},
	pages = {2520--2530},
}

@article{bosi_population-specific_2018,
	title = {Population-specific material properties of the implantation site for transcatheter aortic valve replacement finite element simulations},
	volume = {71},
	issn = {00219290},
	url = {https://linkinghub.elsevier.com/retrieve/pii/S0021929018301143},
	doi = {10.1016/j.jbiomech.2018.02.017},
	language = {en},
	urldate = {2025-08-28},
	journal = {Journal of Biomechanics},
	author = {Bosi, Giorgia M. and Capelli, Claudio and Cheang, Mun Hong and Delahunty, Nicola and Mullen, Michael and Taylor, Andrew M. and Schievano, Silvia},
	month = apr,
	year = {2018},
	pages = {236--244},
}

@article{zhang_sobol_2015,
	title = {Sobol {Sensitivity} {Analysis}: {A} {Tool} to {Guide} the {Development} and {Evaluation} of {Systems} {Pharmacology} {Models}},
	volume = {4},
	issn = {2163-8306},
	shorttitle = {Sobol {Sensitivity} {Analysis}},
	doi = {10.1002/psp4.6},
	language = {eng},
	number = {2},
	journal = {CPT: pharmacometrics \& systems pharmacology},
	author = {Zhang, X.-Y. and Trame, M. N. and Lesko, L. J. and Schmidt, S.},
	month = feb,
	year = {2015},
	pages = {69--79},
}

@article{luraghi_modeling_2019,
	title = {On the {Modeling} of {Patient}-{Specific} {Transcatheter} {Aortic} {Valve} {Replacement}: {A} {Fluid}–{Structure} {Interaction} {Approach}},
	volume = {10},
	issn = {1869-408X, 1869-4098},
	shorttitle = {On the {Modeling} of {Patient}-{Specific} {Transcatheter} {Aortic} {Valve} {Replacement}},
	url = {http://link.springer.com/10.1007/s13239-019-00427-0},
	doi = {10.1007/s13239-019-00427-0},
	language = {en},
	number = {3},
	urldate = {2025-08-28},
	journal = {Cardiovascular Engineering and Technology},
	author = {Luraghi, Giulia and Migliavacca, Francesco and García-González, Alberto and Chiastra, Claudio and Rossi, Alexia and Cao, Davide and Stefanini, Giulio and Rodriguez Matas, Jose Felix},
	month = sep,
	year = {2019},
	pages = {437--455},
}

@article{ghosh_numerical_2020,
	title = {Numerical evaluation of transcatheter aortic valve performance during heart beating and its post-deployment fluid–structure interaction analysis},
	volume = {19},
	issn = {1617-7959, 1617-7940},
	url = {http://link.springer.com/10.1007/s10237-020-01304-9},
	doi = {10.1007/s10237-020-01304-9},
	language = {en},
	number = {5},
	urldate = {2025-08-28},
	journal = {Biomechanics and Modeling in Mechanobiology},
	author = {Ghosh, Ram P. and Marom, Gil and Bianchi, Matteo and D’souza, Karl and Zietak, Wojtek and Bluestein, Danny},
	month = oct,
	year = {2020},
	pages = {1725--1740},
}

@article{basri_fluid_2020,
	title = {Fluid {Structure} {Interaction} on {Paravalvular} {Leakage} of {Transcatheter} {Aortic} {Valve} {Implantation} {Related} to {Aortic} {Stenosis}: {A} {Patient}-{Specific} {Case}},
	volume = {2020},
	copyright = {http://creativecommons.org/licenses/by/4.0/},
	issn = {1748-670X, 1748-6718},
	shorttitle = {Fluid {Structure} {Interaction} on {Paravalvular} {Leakage} of {Transcatheter} {Aortic} {Valve} {Implantation} {Related} to {Aortic} {Stenosis}},
	url = {https://www.hindawi.com/journals/cmmm/2020/9163085/},
	doi = {10.1155/2020/9163085},
	language = {en},
	urldate = {2025-08-28},
	journal = {Computational and Mathematical Methods in Medicine},
	author = {Basri, Adi A. and Zuber, Mohammad and Basri, Ernnie I. and Zakaria, Muhammad S. and Aziz, Ahmad F. A. and Tamagawa, Masaaki and Ahmad, Kamarul A.},
	month = may,
	year = {2020},
	pages = {1--22},
}

@article{fumagalli_fluidstructure_2023,
	title = {Fluid‐structure interaction analysis of transcatheter aortic valve implantation},
	volume = {39},
	issn = {2040-7939, 2040-7947},
	url = {https://onlinelibrary.wiley.com/doi/10.1002/cnm.3704},
	doi = {10.1002/cnm.3704},
	language = {en},
	number = {6},
	urldate = {2025-08-28},
	journal = {International Journal for Numerical Methods in Biomedical Engineering},
	author = {Fumagalli, Ivan and Polidori, Rebecca and Renzi, Francesca and Fusini, Laura and Quarteroni, Alfio and Pontone, Gianluca and Vergara, Christian},
	month = jun,
	year = {2023},
	pages = {e3704},
}

@article{bokma_improved_2023,
	title = {Improved {Outcomes} {After} {Pulmonary} {Valve} {Replacement} in {Repaired} {Tetralogy} of {Fallot}},
	volume = {81},
	issn = {07351097},
	url = {https://linkinghub.elsevier.com/retrieve/pii/S0735109723052324},
	doi = {10.1016/j.jacc.2023.02.052},
	language = {en},
	number = {21},
	urldate = {2025-12-24},
	journal = {Journal of the American College of Cardiology},
	author = {Bokma, Jouke P. and Geva, Tal and Sleeper, Lynn A. and Lee, Ji Hae and Lu, Minmin and Sompolinsky, Tehila and Babu-Narayan, Sonya V. and Wald, Rachel M. and Mulder, Barbara J.M. and Valente, Anne Marie},
	month = may,
	year = {2023},
	pages = {2075--2085},
}

@article{geva_long-term_2024,
	title = {Long-{Term} {Management} of {Right} {Ventricular} {Outflow} {Tract} {Dysfunction} in {Repaired} {Tetralogy} of {Fallot}: {A} {Scientific} {Statement} {From} the {American} {Heart} {Association}},
	volume = {150},
	issn = {0009-7322, 1524-4539},
	shorttitle = {Long-{Term} {Management} of {Right} {Ventricular} {Outflow} {Tract} {Dysfunction} in {Repaired} {Tetralogy} of {Fallot}},
	url = {https://www.ahajournals.org/doi/10.1161/CIR.0000000000001291},
	doi = {10.1161/CIR.0000000000001291},
	language = {en},
	number = {25},
	urldate = {2025-12-24},
	journal = {Circulation},
	author = {Geva, Tal and Wald, Rachel M. and Bucholz, Emily and Cnota, James F. and McElhinney, Doff B. and Mercer-Rosa, Laura M. and Mery, Carlos M. and Miles, Andrea Leann and Moore, Jeremy and {on behalf of the American Heart Association Council on Lifelong Congenital Heart Disease and Heart Health in the Young; Council on Cardiovascular Surgery and Anesthesia; Council on Clinical Cardiology; and Council on Cardiovascular and Stroke Nursing}},
	month = dec,
	year = {2024},
}

@article{lee_long-term_2020,
	title = {Long-term outcomes of pulmonary valve replacement in patients with repaired tetralogy of {Fallot}},
	volume = {58},
	copyright = {https://academic.oup.com/journals/pages/open\_access/funder\_policies/chorus/standard\_publication\_model},
	issn = {1010-7940, 1873-734X},
	url = {https://academic.oup.com/ejcts/article/58/2/246/5733769},
	doi = {10.1093/ejcts/ezaa030},
	language = {en},
	number = {2},
	urldate = {2025-12-24},
	journal = {European Journal of Cardio-Thoracic Surgery},
	author = {Lee, Cheul and Choi, Eun Seok and Lee, Chang-Ha},
	month = aug,
	year = {2020},
	pages = {246--252},
}

@article{geva_repaired_2011,
	title = {Repaired tetralogy of {Fallot}: the roles of cardiovascular magnetic resonance in evaluating pathophysiology and for pulmonary valve replacement decision support},
	volume = {13},
	issn = {10976647},
	shorttitle = {Repaired tetralogy of {Fallot}},
	url = {https://linkinghub.elsevier.com/retrieve/pii/S1097664723013649},
	doi = {10.1186/1532-429X-13-9},
	language = {en},
	number = {1},
	urldate = {2025-12-24},
	journal = {Journal of Cardiovascular Magnetic Resonance},
	author = {Geva, Tal},
	month = jan,
	year = {2011},
	pages = {9},
}

@article{bergersen_harmony_2017,
	title = {Harmony {Feasibility} {Trial}},
	volume = {10},
	issn = {19368798},
	url = {https://linkinghub.elsevier.com/retrieve/pii/S1936879817310257},
	doi = {10.1016/j.jcin.2017.05.034},
	language = {en},
	number = {17},
	urldate = {2025-12-24},
	journal = {JACC: Cardiovascular Interventions},
	author = {Bergersen, Lisa and Benson, Lee N. and Gillespie, Matthew J. and Cheatham, Sharon L. and Crean, Andrew M. and Hor, Kan N. and Horlick, Eric M. and Lung, Te-Hsin and McHenry, Brian T. and Osten, Mark D. and Powell, Andrew J. and Cheatham, John P.},
	month = sep,
	year = {2017},
	pages = {1763--1773},
}

@article{bosi_patientspecific_2015,
	title = {Patient‐specific finite element models to support clinical decisions: {A} lesson learnt from a case study of percutaneous pulmonary valve implantation},
	volume = {86},
	copyright = {http://onlinelibrary.wiley.com/termsAndConditions\#vor},
	issn = {1522-1946, 1522-726X},
	shorttitle = {Patient‐specific finite element models to support clinical decisions},
	url = {https://onlinelibrary.wiley.com/doi/10.1002/ccd.25944},
	doi = {10.1002/ccd.25944},
	language = {en},
	number = {6},
	urldate = {2025-12-24},
	journal = {Catheterization and Cardiovascular Interventions},
	author = {Bosi, Giorgia M. and Capelli, Claudio and Khambadkone, Sachin and Taylor, Andrew M. and Schievano, Silvia},
	month = nov,
	year = {2015},
	pages = {1120--1130},
}

@article{armstrong_association_2019,
	title = {Association between patient age at implant and outcomes after transcatheter pulmonary valve replacement in the multicenter {Melody} valve trials},
	volume = {94},
	issn = {1522-1946, 1522-726X},
	url = {https://onlinelibrary.wiley.com/doi/10.1002/ccd.28454},
	doi = {10.1002/ccd.28454},
	language = {en},
	number = {4},
	urldate = {2025-12-29},
	journal = {Catheterization and Cardiovascular Interventions},
	author = {Armstrong, Aimee K. and Berger, Felix and Jones, Thomas K. and Moore, John W. and Benson, Lee N. and Cheatham, John P. and Turner, Daniel R. and Rhodes, John F. and Vincent, Julie A. and Zellers, Thomas and Lung, Te‐Hsin and Eicken, Andreas and McElhinney, Doff B.},
	month = oct,
	year = {2019},
	pages = {607--617},
}

@article{nordmeyer_acute_2019,
	title = {Acute and midterm outcomes of the post-approval {MELODY} {Registry}: a multicentre registry of transcatheter pulmonary valve implantation},
	volume = {40},
	copyright = {https://academic.oup.com/journals/pages/open\_access/funder\_policies/chorus/standard\_publication\_model},
	issn = {0195-668X, 1522-9645},
	shorttitle = {Acute and midterm outcomes of the post-approval {MELODY} {Registry}},
	url = {https://academic.oup.com/eurheartj/article/40/27/2255/5475832},
	doi = {10.1093/eurheartj/ehz201},
	language = {en},
	number = {27},
	urldate = {2025-12-29},
	journal = {European Heart Journal},
	author = {Nordmeyer, Johannes and Ewert, Peter and Gewillig, Marc and AlJufan, Mansour and Carminati, Mario and Kretschmar, Oliver and Uebing, Anselm and Dähnert, Ingo and Röhle, Robert and Schneider, Heike and Witsenburg, Maarten and Benson, Lee and Gitter, Roland and Bökenkamp, Regina and Mahadevan, Vaikom and Berger, Felix},
	month = jul,
	year = {2019},
	pages = {2255--2264},
}

@article{houeijeh_long-term_2023,
	title = {Long-term outcomes of transcatheter pulmonary valve implantation with melody and {SAPIEN} valves},
	volume = {370},
	issn = {01675273},
	url = {https://linkinghub.elsevier.com/retrieve/pii/S0167527322016527},
	doi = {10.1016/j.ijcard.2022.10.141},
	language = {en},
	urldate = {2025-12-29},
	journal = {International Journal of Cardiology},
	author = {Houeijeh, Ali and Batteux, Clement and Karsenty, Clement and Ramdane, Nassima and Lecerf, Florence and Valdeolmillos, Estibaliz and Lourtet-Hascoet, Julie and Cohen, Sarah and Belli, Emre and Petit, Jérôme and Hascoët, Sébastien},
	month = jan,
	year = {2023},
	pages = {156--166},
}

@misc{maas_sliding_2026,
	title = {Sliding {Interfaces}},
	url = {https://help.febio.org/docs/FEBioUser-4-11/UM411-Subsection-3.14.1.html},
	publisher = {FEBio Theory Manual Version 4.11},
	author = {Maas, Steve A. and Herron, Michael and Weiss, Jeffrey A. and Ateshian, Gerard A.},
	month = jan,
	year = {2026},
}

@article{sturla_planning_2025,
	title = {Planning transcatheter pulmonary valve implantation in the dysfunctional native {RVOT}: {A} semi-automated pipeline for dynamic analysis based on {4D}-{CT} imaging},
	volume = {260},
	issn = {01692607},
	shorttitle = {Planning transcatheter pulmonary valve implantation in the dysfunctional native {RVOT}},
	url = {https://linkinghub.elsevier.com/retrieve/pii/S0169260724005625},
	doi = {10.1016/j.cmpb.2024.108569},
	language = {en},
	urldate = {2026-07-16},
	journal = {Computer Methods and Programs in Biomedicine},
	author = {Sturla, Francesco and Caimi, Alessandro and Giugno, Luca and Pasqualin, Giulia and Tissir, Karima and Secchi, Francesco and Redaelli, Alberto and Carminati, Mario and Votta, Emiliano},
	month = mar,
	year = {2025},
	pages = {108569},
}

@article{pluchinotta_3-dimensional_2020,
	title = {3-{Dimensional} personalized planning for transcatheter pulmonary valve implantation in a dysfunctional right ventricular outflow tract},
	volume = {309},
	issn = {01675273},
	url = {https://linkinghub.elsevier.com/retrieve/pii/S0167527319349253},
	doi = {10.1016/j.ijcard.2019.12.006},
	language = {en},
	urldate = {2026-07-16},
	journal = {International Journal of Cardiology},
	author = {Pluchinotta, Francesca R. and Sturla, Francesco and Caimi, Alessandro and Giugno, Luca and Chessa, Massimo and Giamberti, Alessandro and Votta, Emiliano and Redaelli, Alberto and Carminati, Mario},
	month = jun,
	year = {2020},
	pages = {33--39},
}

@article{astori_fe-based_2025,
	title = {{FE}-based risk assessment of coronary artery compression in pulmonary conduit pre-stenting: optimizing the balance between time-expense and reliability},
	volume = {7},
	issn = {2673-3129},
	shorttitle = {{FE}-based risk assessment of coronary artery compression in pulmonary conduit pre-stenting},
	url = {https://www.frontiersin.org/articles/10.3389/fmedt.2025.1686131/full},
	doi = {10.3389/fmedt.2025.1686131},
	urldate = {2026-07-16},
	journal = {Frontiers in Medical Technology},
	author = {Astori, Davide and Sturla, Francesco and Caimi, Alessandro and Secchi, Francesco and Giugno, Luca and Redaelli, Alberto and Carminati, Mario and Votta, Emiliano},
	month = dec,
	year = {2025},
	pages = {1686131},
}


\end{document}


\renewcommand{\thetable}{A\arabic{table}}
\begin{longtable}{rrrrrr}
\caption{PCE samples of the HGO material model generated for the sensitivity analysis} \label{tab:PCE_Samples} \\
\toprule
\textbf{Sample No.} & \textbf{\(c\)} & \textbf{\(k_1\)} & \textbf{\(k_2\)} & \textbf{\(\kappa\)} & \textbf{\(\gamma\)} \\
\midrule
\endfirsthead

\multicolumn{6}{c}{\tablename\ \thetable\ -- continued} \\
\toprule
\textbf{Sample No.} & \textbf{\(c\)} & \textbf{\(k_1\)} & \textbf{\(k_2\)} & \textbf{\(\kappa\)} & \textbf{\(\gamma\)} \\
\midrule
\endhead

\midrule
\multicolumn{6}{r}{Continued on next page} \\
\endfoot

\bottomrule
\endlastfoot

1  & 117.8815 & 30443.4282 & 0.2695 & 0.0398 & 15.4923 \\
2  & 119.1195 & 7854.9292  & 0.9639 & 0.0482 & 20.0037 \\
3  & 410.8855 & 12544.0386 & 1.2055 & 0.0719 & 10.1554 \\
4  & 684.9904 & 11645.2485 & 0.4053 & 0.0360 & 36.3053 \\
5  & 120.8752 & 14511.9761 & 0.3861 & 0.1541 & 14.8158 \\
6  & 240.1435 & 11949.2607 & 1.7218 & 0.0253 & 17.8082 \\
7  & 134.9608 & 17274.2090 & 0.9915 & 0.0444 & 25.5563 \\
8  & 197.4111 & 11342.3997 & 0.4199 & 0.0610 & 26.8310 \\
9  & 452.2769 & 14300.1813 & 1.6595 & 0.0742 & 25.4984 \\
10 & 529.2126 & 9873.6002  & 1.0334 & 0.1146 & 18.5983 \\
11 & 110.9278 & 11107.6159 & 1.0394 & 0.1085 & 21.9460 \\
12 & 127.0726 & 29633.0386 & 1.9096 & 0.0457 & 23.1182 \\
13 & 158.9222 & 8633.4912  & 1.2719 & 0.0668 & 33.3576 \\
14 & 151.8852 & 21423.7784 & 1.0907 & 0.0604 & 13.2263 \\
15 & 513.7467 & 9379.9780  & 0.4748 & 0.0542 & 14.8116 \\
16 & 184.9228 & 15180.3025 & 1.6143 & 0.1260 & 9.3649  \\
17 & 216.0352 & 6677.0805  & 2.3422 & 0.0747 & 14.4391 \\
18 & 379.0144 & 19801.1740 & 1.2452 & 0.1191 & 14.7889 \\
19 & 94.5369  & 19440.8884 & 3.3634 & 0.0493 & 10.6103 \\
20 & 129.5290 & 23328.3066 & 1.0720 & 0.1428 & 21.1633 \\
21 & 171.6063 & 10922.8554 & 0.8165 & 0.2143 & 28.1256 \\
22 & 712.0693 & 22161.0442 & 0.6606 & 0.1243 & 23.8363 \\
23 & 496.8087 & 25870.1136 & 2.5999 & 0.0782 & 19.3109 \\
24 & 47.2055  & 14293.2536 & 1.1306 & 0.2818 & 11.1811 \\
25 & 188.4394 & 12919.7836 & 1.0474 & 0.0818 & 15.8786 \\
26 & 283.0992 & 34602.1278 & 0.7374 & 0.0695 & 24.7939 \\
27 & 129.2605 & 21133.4345 & 0.7713 & 0.0866 & 18.9708 \\
28 & 108.0030 & 7994.1684  & 1.6454 & 0.1455 & 20.7250 \\
29 & 174.8606 & 14367.4456 & 0.8867 & 0.0829 & 36.8417 \\
30 & 78.2311  & 9580.7234  & 1.5451 & 0.0614 & 13.8020 \\
31 & 102.8183 & 12041.2051 & 1.7499 & 0.0878 & 23.3387 \\
32 & 218.6312 & 11423.8226 & 2.8707 & 0.1371 & 14.8886 \\
33 & 265.8565 & 9647.5266  & 2.1437 & 0.0791 & 29.8529 \\
34 & 538.1825 & 21135.1746 & 0.8638 & 0.2266 & 17.2533 \\
35 & 482.2113 & 6107.7105  & 1.7151 & 0.0799 & 15.3690 \\
36 & 56.8737  & 11425.7105 & 0.5276 & 0.1142 & 12.4264 \\
37 & 110.2420 & 13319.4638 & 0.5806 & 0.0530 & 19.8467 \\
38 & 306.1368 & 11598.2566 & 0.5491 & 0.0465 & 19.3762 \\
39 & 289.8874 & 4397.0704  & 1.2647 & 0.1522 & 25.5727 \\
40 & 227.9252 & 16004.5177 & 1.3953 & 0.0517 & 26.3375 \\
41 & 184.2848 & 14233.8495 & 1.3025 & 0.1279 & 20.0978 \\
42 & 169.4620 & 27475.4694 & 1.2249 & 0.0738 & 12.0750 \\
43 & 178.2133 & 9198.3697  & 0.5875 & 0.1596 & 21.3530 \\
44 & 300.5828 & 10984.3920 & 0.9273 & 0.0513 & 10.1819 \\
45 & 470.5528 & 4405.0698  & 1.6566 & 0.0548 & 32.8595 \\
46 & 205.5287 & 18332.1811 & 1.7791 & 0.0565 & 18.8509 \\
47 & 118.0415 & 19465.7288 & 1.3136 & 0.0893 & 25.9671 \\
48 & 566.8975 & 14280.2009 & 1.7758 & 0.0979 & 11.5914 \\
49 & 304.4363 & 10632.7318 & 1.3514 & 0.1100 & 18.4236 \\
50 & 394.4870 & 10378.1103 & 0.7396 & 0.0797 & 7.7875  \\
51 & 273.8094 & 26190.5038 & 0.5092 & 0.0377 & 13.6861 \\
52 & 74.3642  & 5934.1720  & 0.5576 & 0.0431 & 14.6453 \\
53 & 256.2094 & 16853.1039 & 1.0322 & 0.0804 & 20.0870 \\
54 & 56.9411  & 9116.1668  & 0.3324 & 0.1000 & 19.6703 \\
55 & 422.5957 & 8669.8913  & 2.7418 & 0.0489 & 10.3747 \\
56 & 330.7603 & 18046.9329 & 2.1193 & 0.1062 & 18.6665 \\
57 & 456.6784 & 11660.8911 & 1.3413 & 0.0552 & 21.5892 \\
58 & 326.1465 & 18696.1190 & 0.7530 & 0.0636 & 33.2204 \\
59 & 141.0473 & 15883.9667 & 1.2738 & 0.0989 & 18.3234 \\
60 & 206.5225 & 5476.6378  & 1.9311 & 0.0532 & 25.6419 \\
61 & 62.0492  & 22266.8946 & 1.6999 & 0.2295 & 12.4250 \\
62 & 21.0294  & 16273.9690 & 0.7012 & 0.1534 & 25.6805 \\
63 & 290.1420 & 15636.9773 & 2.4542 & 0.0771 & 22.2277 \\
64 & 100.0435 & 10437.4339 & 1.6613 & 0.1544 & 13.7483 \\
65 & 193.7649 & 6599.5207  & 0.8885 & 0.2425 & 19.8273 \\
66 & 168.1333 & 25908.0195 & 1.2095 & 0.0620 & 17.9509 \\

\end{longtable}